\documentclass[11pt]{article}
\usepackage[a4paper]{geometry}

\usepackage{float}
\usepackage{textcomp}
\usepackage{amsmath}
\usepackage{amssymb}
\usepackage{graphicx}
\usepackage{natbib}
\usepackage{wrapfig}
\usepackage[unicode=true,
 bookmarks=false,
 breaklinks=false,pdfborder={0 0 1},backref=section,colorlinks=true,linkcolor=cyan,citecolor=cyan,urlcolor=blue]
 {hyperref}

\makeatletter

\DeclareTextSymbolDefault{\textquotedbl}{T1}

\usepackage{algorithm}
\usepackage{graphics}
\usepackage{graphicx}
\usepackage{xspace}
\usepackage[dvipsnames]{xcolor}
\usepackage{parskip}
\usepackage{listings}
\usepackage{multicol}
\usepackage{textcomp}

\usepackage[many]{tcolorbox}
\definecolor{main}{HTML}{5989cf}    
\definecolor{sub}{HTML}{cde4ff}     

\newtcolorbox{boxI}{
    colback = sub, 
    colframe = main, 
    boxrule = 0pt, 
    toprule = 6pt 
}
\newcommand{\statslanguage}[1]{{\color{NavyBlue}#1}}

\renewcommand{\section}{\@startsection%
{section}{1}{0mm}{-\baselineskip}%
{0.5\baselineskip}{\normalfont\Large\bfseries}}%

\usepackage{enumitem}
\setlist[itemize]{align=parleft,left=0pt..1em}

\newlength{\remaining}

\usepackage{fancyhdr}
\date{}
\makeatother

\begin{document}
\title{
Statistical Aspects of X-ray Spectral Analysis}
\author{Johannes Buchner*$^{1}$ \& Peter Boorman$^{2, 3}$}
\maketitle

\begin{center}
*Corresponding author

$^{1}$\textit{Max Planck Institute for Extraterrestrial Physics, Giessenbachstrasse, 85741 Garching, Germany}
\url{jbuchner@mpe.mpg.de}

$^{2}$\textit{Astronomical Institute of the Czech Academy of Sciences, Bo\v{c}n\'i II 1401, CZ-14100 Prague, Czech Republic}

$^{3}$\textit{Cahill Center for Astrophysics, California Institute of Technology, 1216 East California Boulevard, Pasadena, CA 91125, USA}
\end{center}

\vspace{1cm}

Let's imagine that we have received an X-ray spectral file with corresponding auxiliary files (.arf and .rmf)\footnote{The spectral files for the exercises in this textbook chapter can be downloaded from \url{https://github.com/pboorm/xray_spectral_fitting}.}, known to come from an observation of a point source. We are told by our advisor to ``fit the spectrum with a power law''. Section~\ref{sec1} explores what that means exactly. There are subtleties involved when inferring physically-meaningful information about a system from an X-ray spectrum, which can seem daunting to even experienced X-ray astronomers. This chapter 
provides a step-by-step introduction (Section~\ref{sec1}) to performing X-ray spectral fitting in practice, and investigates the subtleties involved. 
Frequentist data analysis and Bayesian inference are discussed in Sections~\ref{sec2} and~\ref{sec3}, respectively. Hands-on exercises are included for both. Section~\ref{sec4} and~\ref{sec5} cover more advanced concepts, such as a framework for inferring sample distributions. Further reading material for specialised and advanced topics is listed in Section~\ref{sec5}.

To cover the essential concepts, the scope of this chapter is necessarily limited. Mathematical details on statistics can be found in the further reading section at the end (Section \ref{sec5}). The examples and discussion focus on the analysis of an isolated X-ray point source observed with focusing optics and a charge-coupled detector. From this case we hope the reader can apply the learned concepts to other situations. The hands-on exercises focus on two widely used X-ray spectral analysis packages, Sherpa and Xspec. However, the same analyses could be made with other packages as well.

\section{The story of detected X-ray photon counts}\label{sec1}

\begin{figure}
\begin{centering}
\includegraphics[width=0.99\textwidth]{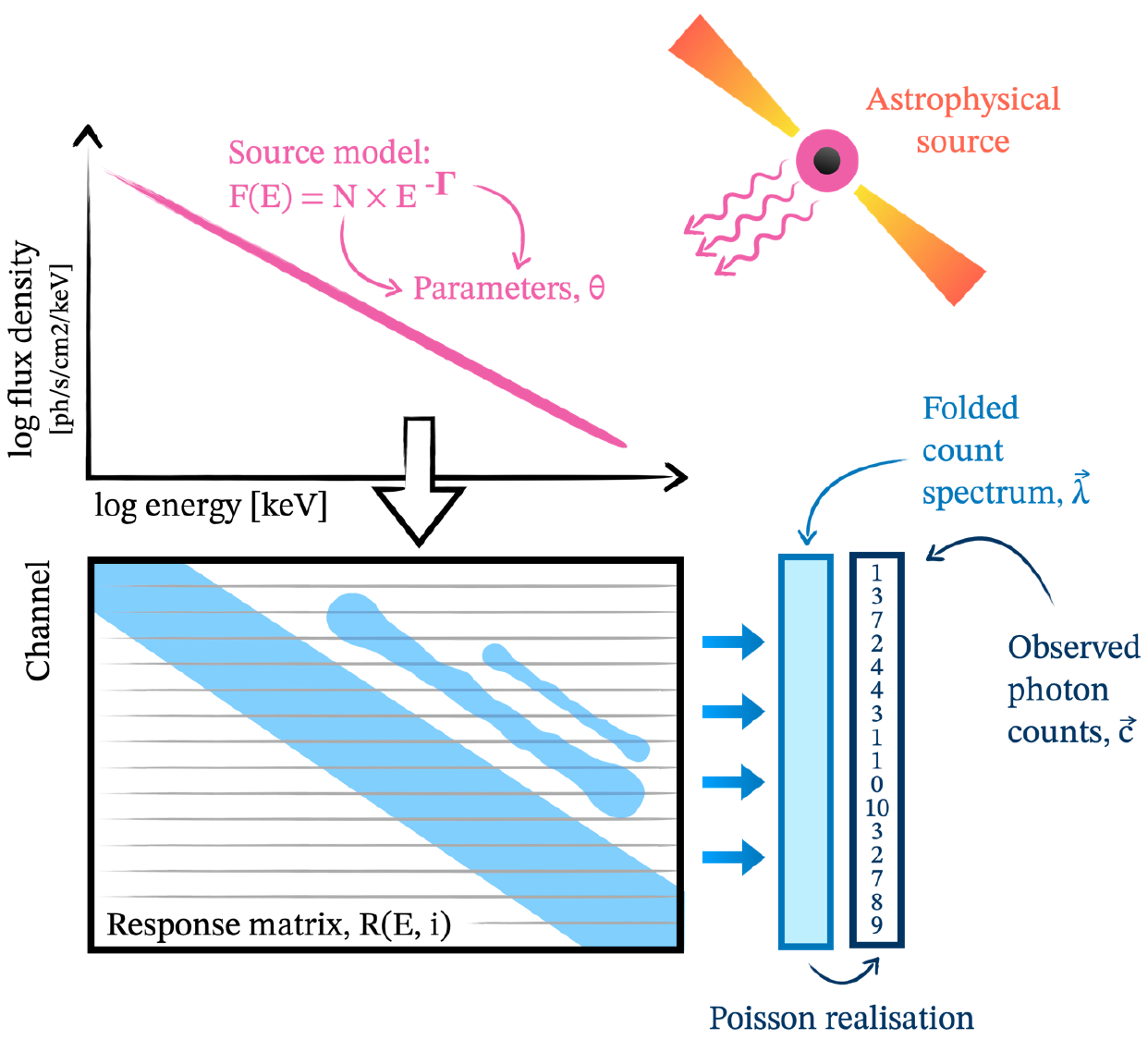}
\par\end{centering}
\caption{\textbf{The X-ray measurement process and its linear approximation (eq.~\ref{eq:linapprox}).} The X-ray spectrum emitted by the astrophysical source (pink, $F(E)$) is modified by the instrument response (matrix), which encodes the energy-dependent response. The observed X-ray events are recorded in detector channels (horizontal lines).
The spectral model $F(E)$ is forward folded through a response matrix
$R$ to obtain a folded count spectrum $\vec \lambda$.
The observed photon counts $\vec c$ are a Poisson realisation.\label{fig:linear}}
\end{figure}

X-ray observations are typically based on photon counting, where each count has a reconstructed energy which is not trivially related to the original X-ray photon's energy.
As discussed in \citet{Fioretti20}, the process of collecting X-ray photon counts in the case of spectro-imaging detectors behind
Wolter-type optics without pile-up can be linearly approximated. Then in a detector channel $i$, the expected number of detected count events is:
\begin{align}
\lambda_{i} & =\int_{0}^{\infty}R_{i}(E)\times F(E)\times\Delta t\,dE\label{eq:linapprox}\\
 & \approx\sum_{j}R_{ij}\times A_{j}\times F_{j}\times\Delta t\times\Delta E_{j}
\end{align}
Here $E$ is the photon energy in keV, $\Delta t$ is the exposure
time. The time-averaged spectral flux density of the source
of radiation in units of erg/s/cm\texttwosuperior /keV, $F(E)$ is represented by a vector $F_j$. The X-ray measurement process is shown in Figure~\ref{fig:linear}.

\begin{figure}
\begin{centering}
\includegraphics[width=0.99\textwidth]{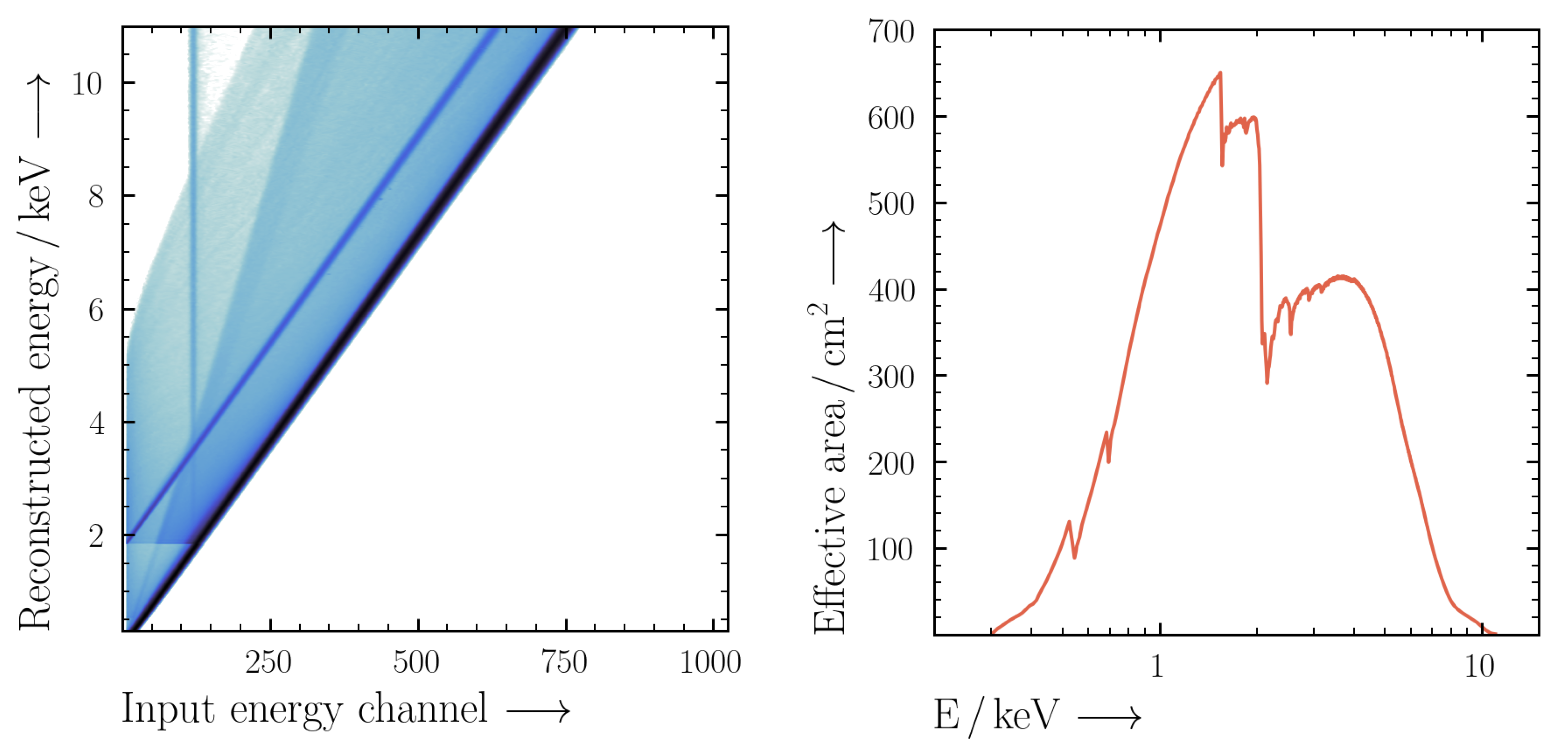}
\par\end{centering}
\caption{\textbf{Response files for an X-ray spectrum.} \textit{Left panel:} An example energy redistribution matrix (.rmf file) for a \textit{Chandra}/ACIS observation. A given energy channel detects a charge, which has a strong probability of having originated from a particular energy photon (the dark diagonal band). However, there is still a finite chance that the charge detected was excited by a photon of higher energy that triggered secondary photons. \textit{Right panel:} The effective area (.arf file) for the same observation, describing the telescope's sensitivity at particular energies. The strong absorption edges are caused by incoming photons interacting with the telescope mirrors.\label{fig:rmf_arf}}
\end{figure}

The instrument response, $R$, can be separated into a row-wise normalised
energy redistribution matrix $R_{ij}$ and a normalising area $A_{j}$,
which carries the units of $\mathrm{cm}{{}^2}$ and describes the
telescope's sensitivity at that energy (see Figure~\ref{fig:rmf_arf}). $R_{ij}$ is typically not
invertible, so we cannot infer $F$ uniquely from the spectrum. Instead we need to forward-fold the measurement process to identify plausible physical spectra $F$ that match the observed data. The linear approximation is illustrated in Figure~\ref{fig:linear}.
Nevertheless, $R$ often has a diagonal band, which allows assigning
the detector channels a nominal energy designation, which is helpful
for interpretation and visualisation.

\begin{figure}
\begin{centering}
\includegraphics[width=0.99\textwidth]{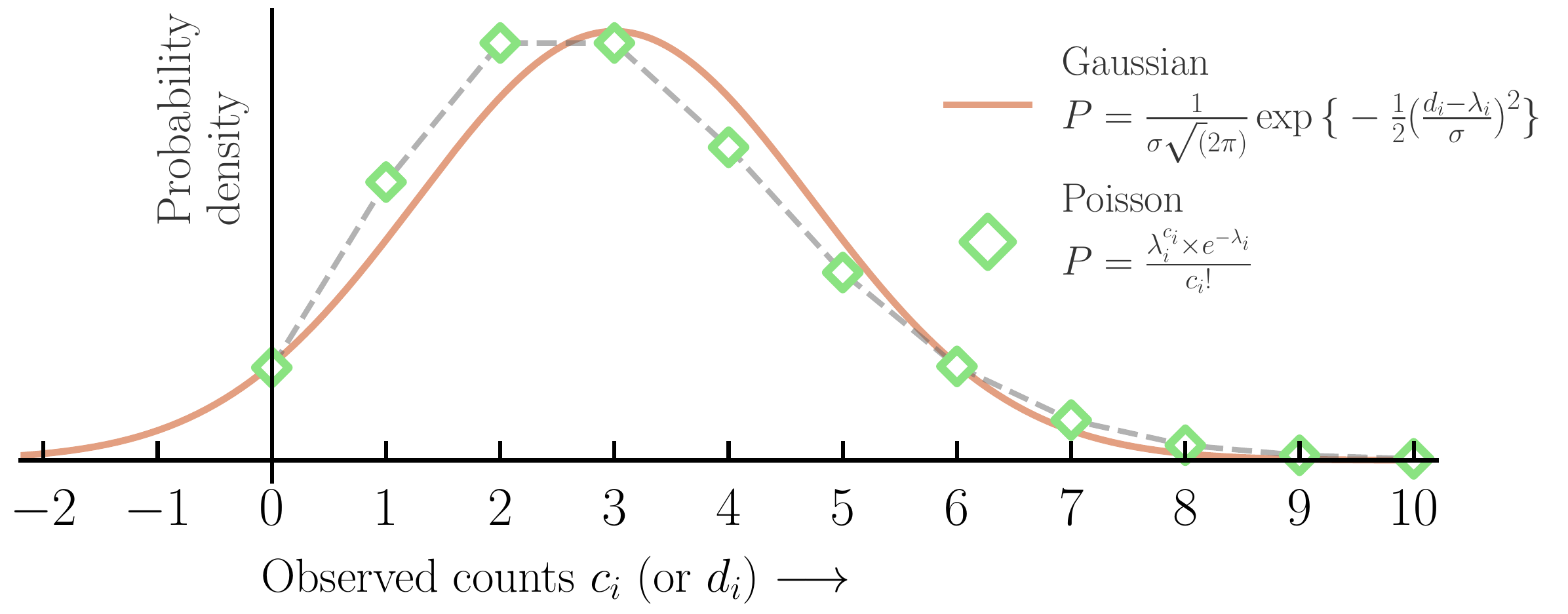}
\par\end{centering}
\caption{\textbf{The Poisson and Gaussian probability distributions.} There are six key differences: Poisson is defined over (1) non-negative (2) integer values, while the Gauss can be negative and is
continuous. (3) The Gauss distribution is symmetric,
while the Poisson is asymmetric. (4) At the low value end the Poisson
distribution is steeper than the Gauss, while (5) at the high end, the
Poisson is shallower than the Gauss. (6) The Gauss distribution has two
parameters (mean and standard deviation), while the Poisson has one
(expected number of counts).\label{fig:poissonshape}}
\end{figure}

Under the linear approximation, given an astrophysical source model ($F$),
and an instrument model ($R$), we can predict the average number of
photon counts detected, $\lambda_{i}$ in one detector energy channel.
Say, for example, $\lambda_{i}=4.2$ counts are expected. Now if you
look into your spectral fits file, you will see that the number of
counts in each channel, $c_{i}$, is an integer number (0, 1, 2, etc),
which describes how many photon events have been detected and assigned
to a given energy channel. The number of counts assigned to each energy channel is therefore a stochastic realisation
of the expected number $\lambda_{i}$, and can be described by a Poisson
process:

\begin{equation}
c_{i}\sim\mathrm{Poisson}(\lambda_{i})
\end{equation}
In words: $c_{i}$ is randomly sampled from a Poisson distribution
with expectation $\lambda_{i}$. The Poisson probability to sample
$c_{i}$ counts given an expected number of counts of $\lambda_{i}$
is (illustrated in Figure~\ref{fig:poissonshape}):

\begin{equation}
P_{\mathrm{Poisson}}(c_{i},\lambda_{i})=\frac{\lambda_{i}^{c_{i}}\times e^{-\lambda_{i}}}{c_{i}!}\label{eq:poissonpdf}
\end{equation}
The likelihood describes how frequently, under the assumed model
$\lambda(R,F,...)$, we expect to see $c_{i}$ counts. This bears
repeating: it describes the tendency of a given model to produce $c_{i}$
counts.

This is convenient since we can now try to search for models which have a
high probability of producing the data (i.e. counts) that we observe! But first there are a few things to tighten up.

\subsection{Combining independent data}

Since we have data in multiple energy channels (it is a spectrum,
after all), we want to use them. Since the probabilities are independent,
the probability of all data is the product of the individual ones:

\begin{equation}
{\cal L}(\theta)=P(\overrightarrow{c},\overrightarrow{\lambda}(\theta))=\prod_{i}P(c_{i},\lambda_{i}(\theta))\label{eq:probprod}
\end{equation}
Given the prediction $\lambda_{i}$ in each channel, which depends
on the model parameters $\theta$, this likelihood ${\cal L}(\theta)$
gives us the probability over all data counts considered.

Eq.~\ref{eq:probprod} also applies for simultaneously analysing multiple data sets. For example, when combining data from a low-energy instrument such as \textit{eROSITA} with a high-energy instrument such as \textit{NuSTAR}. In each considered detector channel the predictions are compared to the observed counts and the probabilities multiplied following eq.~\ref{eq:probprod}.

\begin{figure}
\begin{centering}
\includegraphics[width=0.99\textwidth]{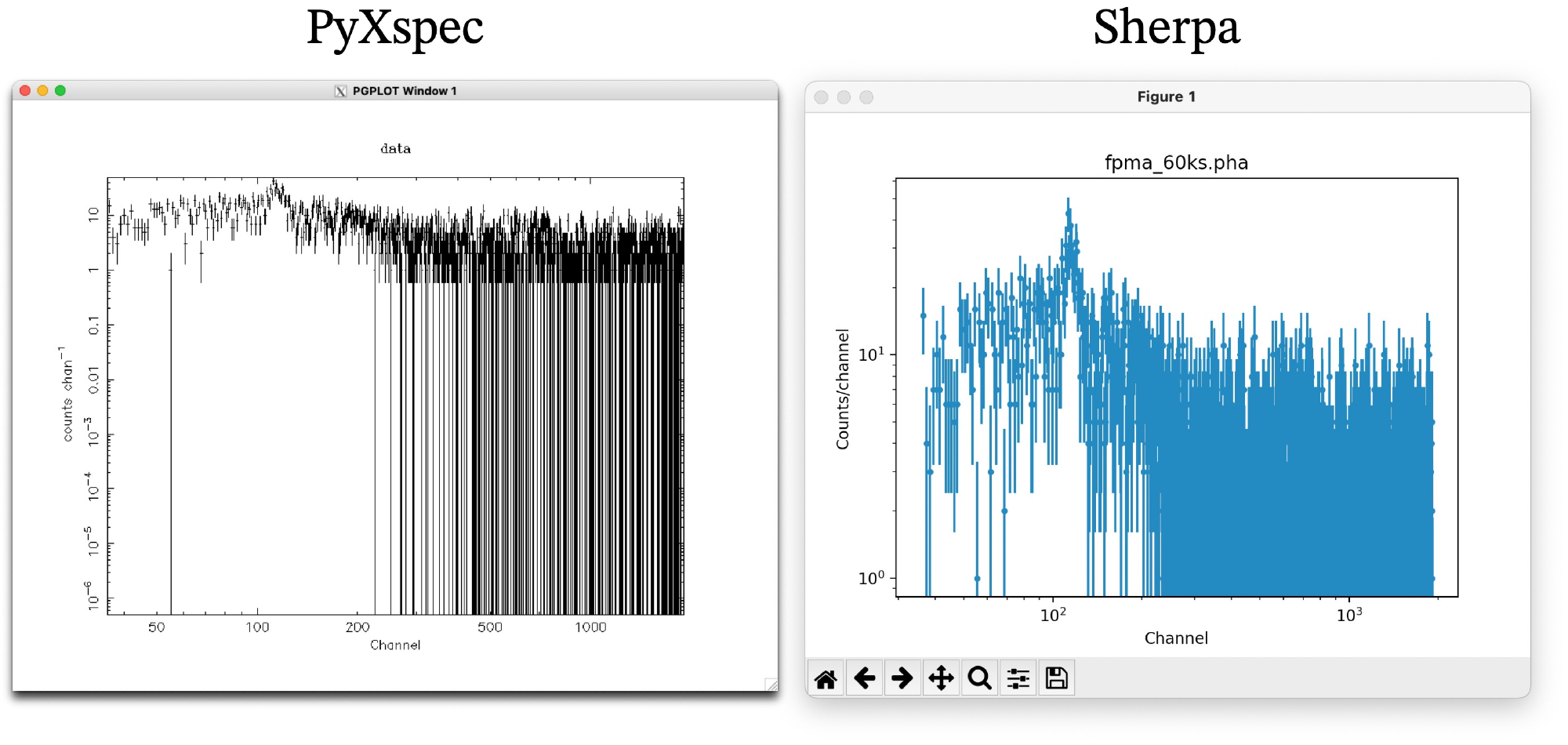}
\par\end{centering}
\caption{Detected channel vs. number of counts for Exercise~1 with PyXspec (left) and Sherpa (right). The error bars with long tails to the bottom are those with one count, where the error is estimated to be also 1, thus reaching down to zero. Data points corresponding to zero counts are missing altogether in this visualisation.\label{fig:ex1}}

\end{figure}

\begin{boxI}
\subsubsection*{Exercise 1 -- Meet the data}
Now let's test this first hand with the spectrum that was provided to us at the start of the Chapter. First, we load the data into PyXspec and/or Sherpa and produce an unbinned plot of detected channel vs. number of counts for the on region used for extracting counts (i.e. in which the background counts are present). Note that for these examples, the response file (.rmf), effective area file (.arf) and background file have been tied in the spectrum's fits header so that PyXspec and Sherpa know to load them automatically with the spectrum. If these have not been set, you can set them manually by updating the keywords in the header of the spectrum or by using the \texttt{grpppha} \texttt{ftool} commands \texttt{chkey RESPFILE}, \texttt{chkey ANCRFILE}, \texttt{chkey BACKFILE} for the response, effective area and background respectively.\\

\textbf{Sherpa}:\\
(First activate your \texttt{CIAO} environment and start Sherpa interactively by typing \texttt{sherpa} to start a Sherpa iPython session).
\begin{lstlisting}[frame=single]
load_pha(1, "fpma_60ks.pha")
ignore_id(1, "0.:3.,78.:")
set_analysis(1, "chan", "counts")
plot_data(xlog=True, ylog=True)
\end{lstlisting}

\textbf{PyXSpec}:\\
(First initialise HEASOFT and start an iPython session).
\begin{lstlisting}[frame=single]
from xspec import *
Plot.xAxis = "chan"
Plot.device = "/xw"
AllData("1:1 fpma_60ks.pha")
AllData.ignore("1: 0.-3. 78.-**")
AllData(1).background = None
Plot.addCommand("log x on")
Plot("lcounts")
\end{lstlisting}

The plots that are produced are shown in Figure~\ref{fig:ex1}.

\end{boxI}

\subsection{Understanding chi\texttwosuperior{} and CStat}

Now we have the machinery set up to calculate the probability
of the process to make the observed data, given a source flux model
and its parameters $\theta$. The natural next step is to try to find
the model parameters that maximize this probability. It has become
convention in computer science to build machinery that minimizes (cost) functions, which can be
adopted by flipping the sign of a function to maximize. For numerical accuracy it is convenient to work in logarithms. Taken together, it has become
standard practice to minimize the fit statistic $-2\times\log P(\theta)$.

\begin{figure}
\begin{centering}
\includegraphics[width=0.8\textwidth]{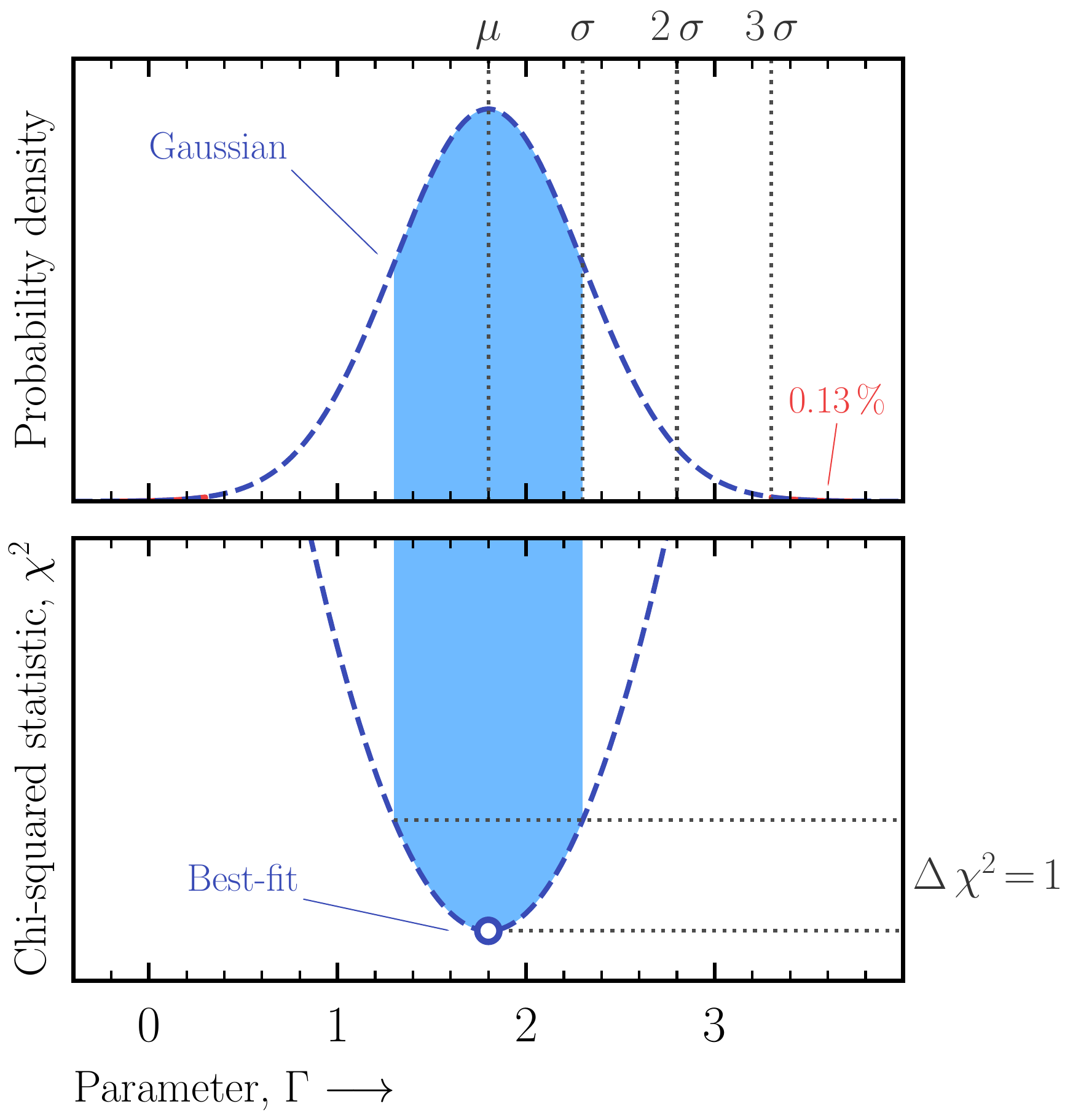}
\par\end{centering}
\caption{\textbf{Relation between Gaussian distribution and fit statistic `$\chi^{2}$'.} The Gaussian distribution has mean $\mu$ and standard
deviation $\sigma$, whereas `$\chi^{2}$' is twice the negative
logarithm. The area within 1 sigma ($\Delta\chi^{2}<1$) and beyond
$3\sigma$ is illustrated. \label{fig:chi2}}

\end{figure}

For the Poisson likelihood (eq.~\ref{eq:poissonpdf}), and dropping
constants, we obtain the statistic:
\begin{equation}
\mathrm{``CStat"}=\sum_{i}c_{i}\log\lambda_{i}-\lambda_{i} \label{eq:cstat}
\end{equation}
If one obtained instead channel data drawn from a Gaussian process,
\begin{equation}
d_{i}\sim\mathrm{Normal}(\lambda_{i},\sigma_{i})
\end{equation}
then the likelihood is 
\begin{equation}
P_{\mathrm{Normal}}(d_{i},\lambda_{i})=\frac{1}{\sqrt{2\pi}\sigma}\exp\left\{ -\frac{1}{2}\left(\frac{d_{i}-\lambda_{i}}{\sigma}\right)^{2}\right\} 
\end{equation}
and the minimization statistic, following again $-2\times\log P(\theta)$,
becomes:

\begin{equation}
``\chi^{2}"=\sum_{i}\left(\frac{d_{i}-\lambda_{i}}{\sigma_{i}}\right)^{2}
\label{eq:chi2}
\end{equation}
Here we see where the multiplication of two came from. It makes the
form for the Gaussian case most convenient. The correspondence of
$\chi^{2}$ and Gaussian probability density is illustrated in Figure~\ref{fig:chi2}.

Thus saying ``I use chi\texttwosuperior{} statistics'' is a short-hand
for ``I assume a Gaussian measurement model'', while ``I
use CStat'' is shorthand for ``I assume a Poisson count process''.
The former has an additional uncertainty parameter for each channel,
$\sigma_{i}$.

\begin{wrapfigure}{r}{0.5\textwidth}
\begin{centering}
\vspace{-0.5cm}
\includegraphics[width=0.49\textwidth]{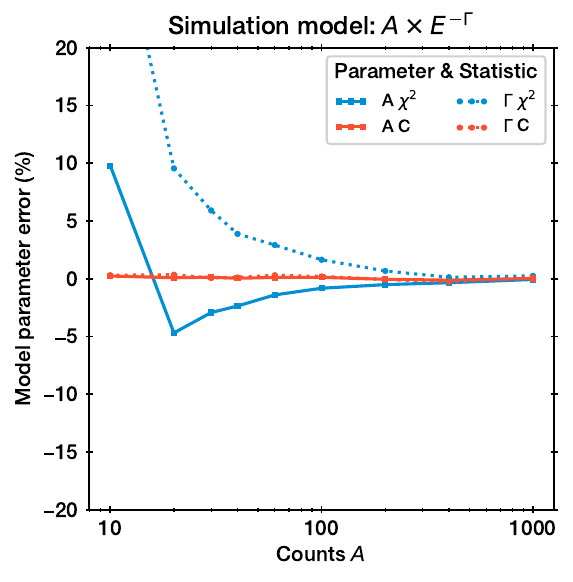}
\vspace{-0.5cm}
\par\end{centering}
\caption{A $\chi^2$-based fit (blue curves) of a power-law under-estimate the amplitude (solid curve) and over-estimate the powerlaw index (dotted curve). C-stat (green) is unbiased. 
Based on \cite{Mighell99}.
\label{fig:poissondemos}
}
\vspace{-0.75cm}
\end{wrapfigure}

When should one employ C-Stat and at what point is $\chi^2$, with some $\sigma_i$, an appropriate approximation? The $\chi^2$ statistic leads to biased model parameter estimates, as illustrated in Figure~\ref{fig:poissondemos}. Poisson data were simulated from a powerlaw model $A\times i^{-\Gamma}$  where i is the channel index from 0 to 100, $A$ is the powerlaw amplitude and $\Gamma=1$ is the powerlaw index. Then, the model parameters were fit for by optimization. $\chi^2$ statistics are biased by over 5 per cent when using fewer than $\sim$40 counts, with the amplitude under- and the index over-estimated. This bias decreases towards higher counts. However, a systematic offset remains that is comparable to the statistical uncertainties and thus has a statistically significant impact at all counts \citep{Humphrey2009}.

In contrast, C-Stat is unbiased at all counts \citep{Cash79}.  Figure~\ref{fig:poissondemos} also shows that switching from C-Stat to $\chi^2$ when the spectrum has at least 30 counts causes a 5 per cent discontinuity in the estimated model parameters.
Historically, faster computation time favoured $\chi^2$, however, today, computation is dominated by source models rather than the statistic. Fast computers and modern inference algorithms mean C-Stat can be used at all levels of count statistics. More information can be found in \cite{Wheaton1995,Nousek1989,Mighell99} and the introduction of \cite{Dyk2001}.



\subsection{Detector details, binning and grouping}
When an X-ray hits a charge-coupled detector, an electron charge is deposited onto one or more pixels. These are then grouped into an event, and the electron charge estimated (see \citealt{Fioretti20} for more details). The charge is converted into an energy channel, on a binning that makes sense for the given instrument.
The binning can sometimes be chosen by the user of the analysis pipeline, or modified after. 

The binning is often rather fine, to not hinder any scientific investigations.
However, the detector response sets some fundamental limits on how much information, even with high numbers of counts, can be extracted. \cite{Kaastra2016OptimalBinning} proposed optimal binning of the spectrum based on the detector response. This typically reduces the computation time in parameter estimation, with essentially no loss of information. We recommend this procedure, which is implemented in \texttt{ftgrouppha}\footnote{\url{https://heasarc.gsfc.nasa.gov/lheasoft/help/ftgrouppha.html}} (included with HEASARC's ftools; see exercise 3). While this re-binning method can be considered a good starting point, some fit statistics may require further, coarser re-binning to reach a minimal number of counts.

A further motivation to apply coarser binning is that the detector response is not perfectly describing the instrument behaviour. For example, detector edges or peaks may not be positioned perfectly. Such systematics can bias the spectral fit. 

To reach the regime where $\chi^2$-based estimates are approximately valid (see Figure~\ref{fig:poissondemos}), further grouping of the detector channels can be applied as a data pre-processing step. Strategies include binning n-fold, adaptively by signal-to-noise ratio or minimum number of counts, by tools such as \texttt{ftgrouppha} and Sherpa\footnote{\url{https://cxc.cfa.harvard.edu/sherpa/threads/pha_regroup/}}. Broad energy bands (e.g., 0.5-2\,keV, 2-10\,keV, 15-195\,keV) are an extreme case. In the process of such rebinning, spectral shape information is always lost. How much information is lost depends on the spectral model. In case of smooth continuum models, the impact may be less severe than for identifying narrow lines. Rebinning is unnecessary when Poisson statistics are used. Bins with zero counts are fine. However, see the next section for how handling background spectra change the situation.

In addition to the requirements for fitting, it can be difficult to visualise the shape of the observed spectrum without some level of binning. Therefore, separately from the statistical analysis, it is useful to rebin the data for visualisation purposes. Within Xspec, the \texttt{setplot rebin} command (\texttt{Plot.setRebin()} in PyXspec), combines adjacent bins to have some minimum significance, which can make it easier to interpret the shape of the data being fit. This command has no effect on the fitting.

\subsection{Background spectra}
\begin{figure}
\begin{centering}
\includegraphics[width=0.99\textwidth]{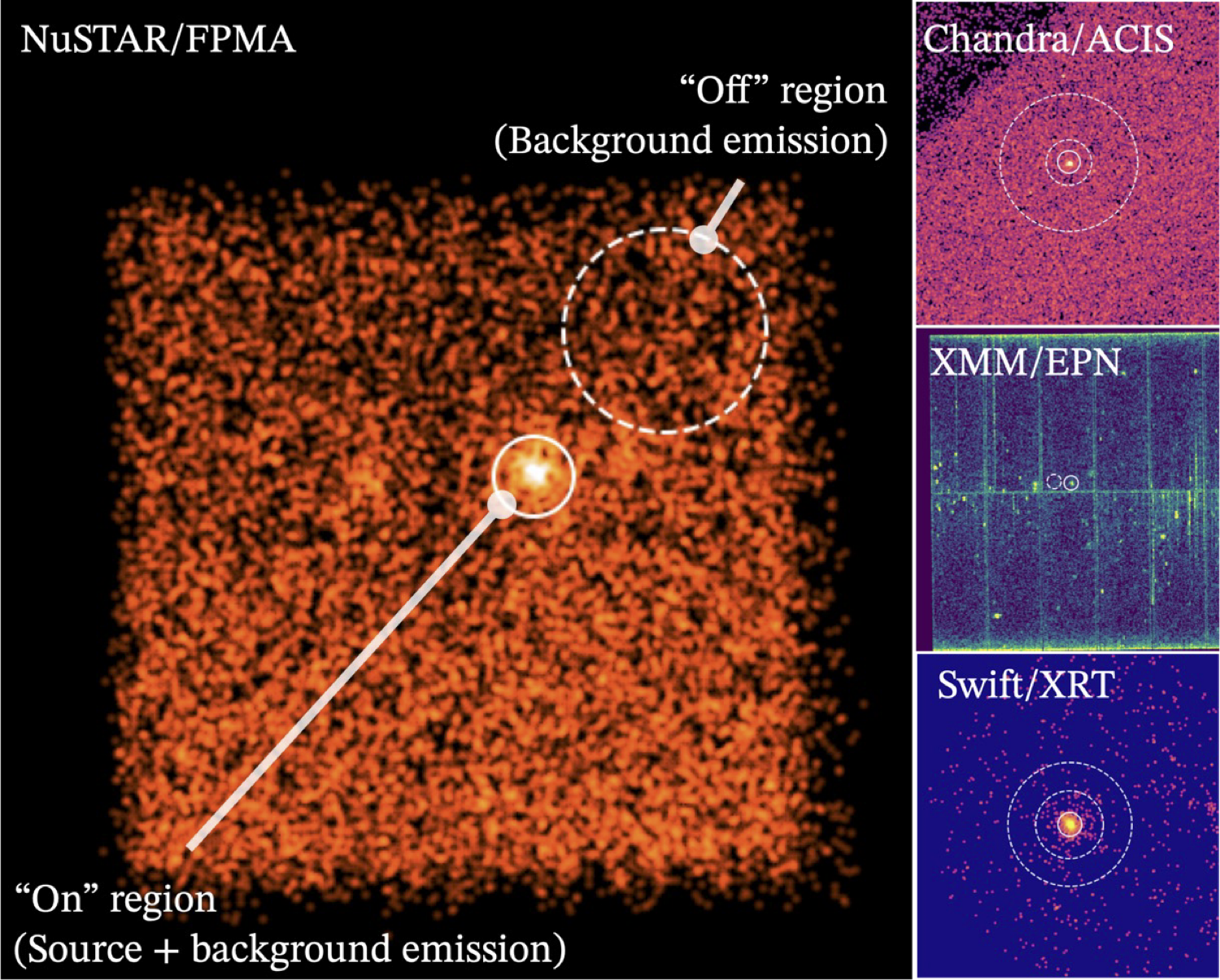}
\par\end{centering}
\caption{\textbf{On and off region treatment.} \label{fig:onoff} In the typically larger off/background region (dashed lines in each panel), only the background processes are assumed to contribute. In the on/source region (solid lines), both the background and source emission contribute. Examples are shown for \textit{NuSTAR}/FPMA, \textit{Chandra}/ACIS, \textit{XMM-Newton}/EPN and \textit{Swift}/XRT.}
\end{figure}

A special case of the combination of data is when one spectrum was taken
in an ``off'' time or ``background region'', where only background
radiation processes are assumed to contribute, and another spectrum was taken in an ``on'' time or
``source region'', where both the background and the source process
of interest contribute. This is illustrated in Figure~\ref{fig:onoff}.
%
%
%

\begin{wrapfigure}{r}{0.5\textwidth}
\begin{centering}
\vspace{-0.8cm}
\includegraphics[width=0.49\textwidth]{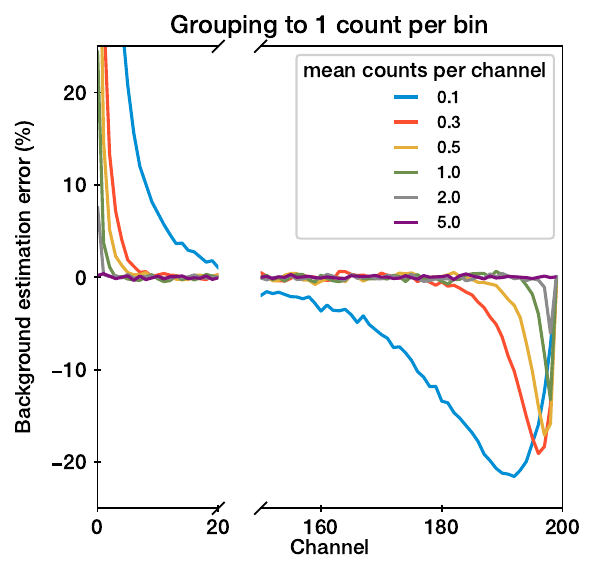}
\vspace{-0.5cm}
\par\end{centering}
\caption{When binning the background to at least one count per group, the background level in the first (last) channels tend to be systematically over- (under-)estimated. This is caused by the rebinning adaptively changing the bin edges starting from the lowest filled channel and ending at the high end, which causes fluctuations of the bin edges and widths.
\label{fig:wstatdemo}
}
\vspace{-0.5cm}
\end{wrapfigure}

To use eq.~\ref{eq:probprod} in this case, we have for the
 on region $\lambda_{i}^{\mathrm{on}}=\lambda_{\mathrm{src},i}+\lambda_{\mathrm{bkg},i}$
and for the off/background region $\lambda_{i}^{\mathrm{off}}=\lambda_{\mathrm{bkg},i}$.
In the process of computing $\lambda$ from the flux, different areas and exposure times for the on and off regions may need to be
 considered. 

In some cases, the background contribution to the observed counts is much smaller than the source contribution at all considered energy channels, and it can be ignored. Otherwise, a model for the background needs to be defined. These can be either non-parametric or parametric, and informed from other observations or solely based on the observation at hand.

\begin{figure}
\begin{centering}
\includegraphics[width=0.99\textwidth]{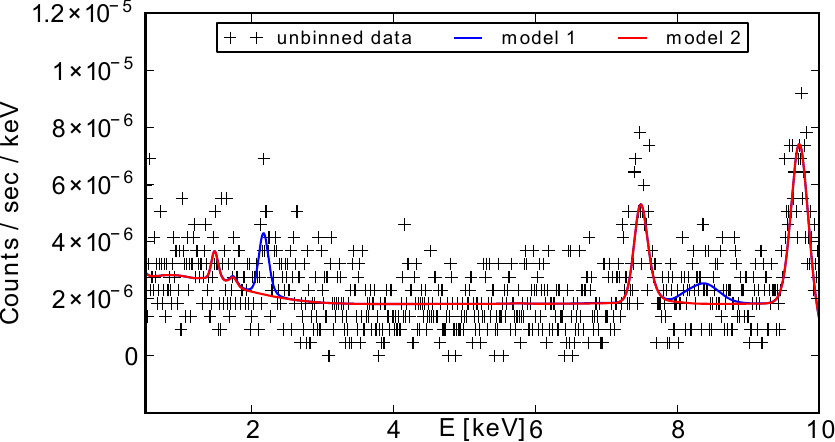}
\par\end{centering}
\caption{\textbf{Chandra example background spectrum.} \label{fig:bkg} The spectrum
is divided by exposure time and bin energy width. Still, the Poisson
nature (non-negative integer values) is clearly visible. Two empirical
background models of different complexity consisting of continuum
and line shapes are plotted in blue and red. Taken from \cite{Buchner2014}.}
\end{figure}

Figure~\ref{fig:bkg} shows an example of an auxiliary background model fitted to a Chandra background. The model is an empirical mixture of power laws and Gaussians, without attempting to assign physical meaning to the components.
For other examples of background modelling, see \cite{Wik14NuSTARbkg} (physical model for NuSTAR\footnote{See \url{https://github.com/NuSTAR/nuskybgd-py}}) and \cite{Maggi14XMM} (semi-physical model for XMM-Newton).
Defining such models can have the benefit of (1) propagating the background uncertainties
and (2) taking into consideration that the background may not vary
rapidly between detector channels. The latter is reasonable when there
are no strong response edges or in the case of the cosmic particle
background. Parametric background models can also be machine-learned by analysing large sets of archival observations \citep[see Appendix of][]{Simmonds18XZ}, which do not necessitate physical knowledge of the background itself.

Sometimes we are not able to build a detailed model for the background. In that
case, we could try to estimate the background contribution in each
detector channel directly. A naive estimate with the channel counts of the background region would be $\lambda_{\mathrm{bkg,i}}^{\mathrm{off}}\approx b_{i}$. A less biased estimate can be obtained by maximum likelihood. Then we can add $\lambda_{\mathrm{bkg,i}}$ to the
source count prediction. Taking into consideration differences
in region definition, that is: 
\begin{align}
\lambda_{i}^{\mathrm{on}}=\lambda_{\mathrm{src},i}+\lambda_{\mathrm{bkg,i}}^{\mathrm{off}}\times\frac{A_{\mathrm{on}}}{A_{\mathrm{off}}}\times\frac{\Delta t_{\mathrm{on}}}{\Delta t_{\mathrm{off}}}
\end{align}
This profile likelihood (meaning the optimum for the nuisance parameter was determined) approach is called the WStat statistic \citep{Wachter79}, and is a modification
of CStat \citep{Cash79}.
Instead of subtracting the background, which would not depart from Poisson statistics (Skellam statistics), WStat adds a background contribution to the source model in each channel.


Fitting with WStat has two complications: firstly, the uncertainty by not knowing $\lambda_{\mathrm{bkg,i}}^{\mathrm{off}}$
precisely is not propagated into the source flux estimate. Secondly, biases in the predicted source counts can occur if the background is not sampled
with a sufficient number of counts, in a sufficient fraction of the background bins.

In such cases, some bins will
have zero counts, the background will be underestimated and the source
flux will be over-estimated. 
These issue are explored in a numerical study\footnote{\url{https://giacomov.github.io/Bias-in-profile-poisson-likelihood/}\label{giacomoprofilelikelihood}} by Giacomo Vianello. 

One way to rectify this problem is to adaptively bin the source spectrum such that the background spectrum bins contain a minimum number of counts. In general, data-dependent rebinning, and then estimating something from the same data, should be treated with caution. Indeed, rebinning to a minimum of 1 count per bin is often biased, as illustrated in Figure~\ref{fig:wstatdemo}. The numerical study\textsuperscript{\ref{giacomoprofilelikelihood}} shows that grouping to a minimum of 5 bins seems to give an acceptably low bias. The grouping needs to be based on the counts in the background spectrum. 
The grouping needs to be stored in the source spectrum (see exercise~3), because the source spectrum binning determines how the background spectrum is treated, in e.g., Xspec.
The discussed issues are relevant for WStat statistics, which is used by default in Xspec when the statistic is set to "cstat" and a background is loaded.

\begin{figure}
\begin{centering}
\includegraphics[width=0.99\textwidth]{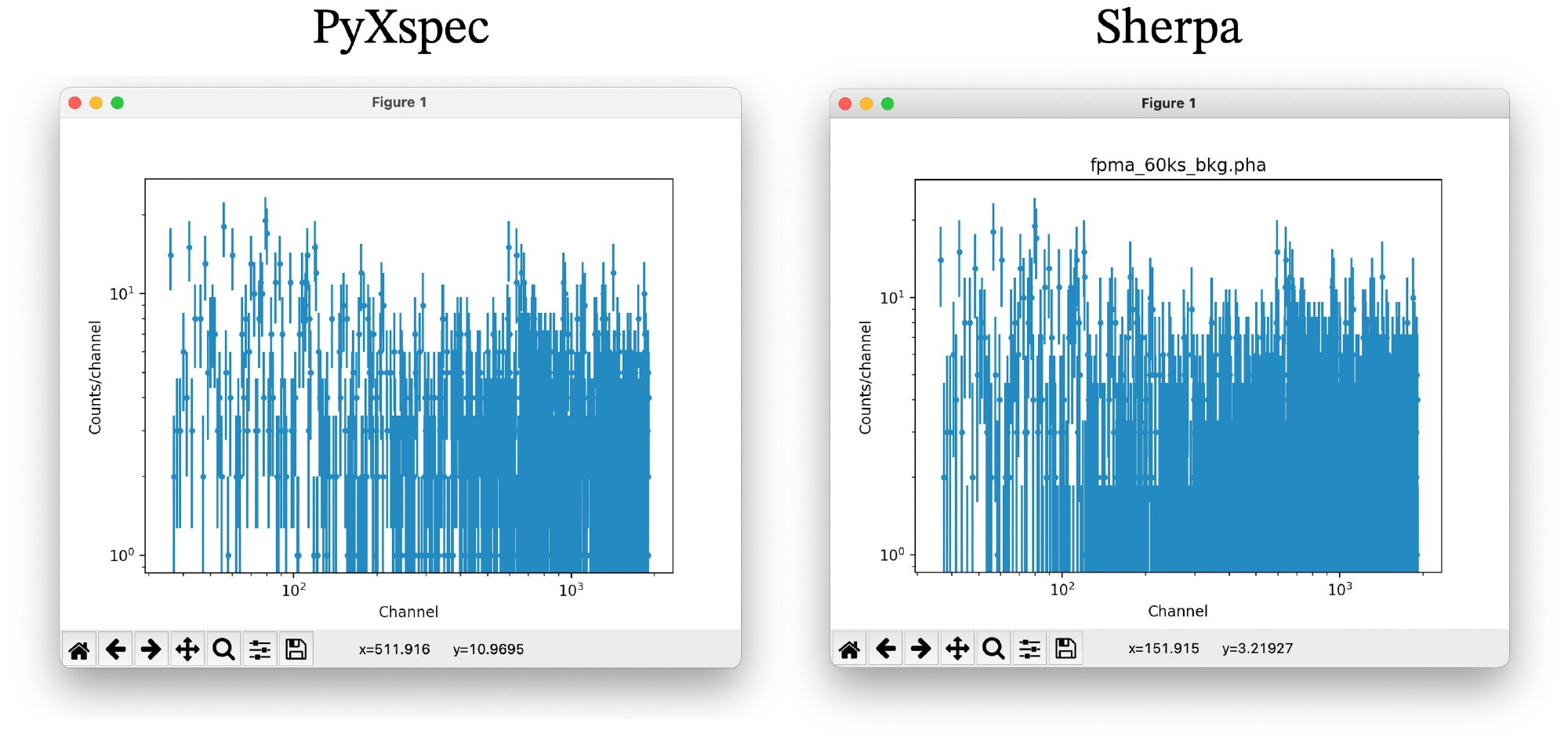}
\par\end{centering}
\caption{Detector channel vs. number of background counts for Exercise~2 with PyXspec (left) and Sherpa (right). \label{fig:ex2}}

\end{figure}

\begin{boxI}
\subsubsection*{Exercise 2 -- visualising the background}
Following on from Exercise~1, we plot the background counts in our observation. \\

\textbf{Sherpa}:
\begin{lstlisting}[frame=single]
plot_bkg(xlog=True, ylog=True)
\end{lstlisting}

\textbf{PyXSpec}:
\begin{lstlisting}[frame=single]
import numpy as np
import matplotlib.pyplot as plt
Plot.background = True
Plot("lcounts")
y = np.array(Plot.backgroundVals())
x = Plot.x()
xErr = Plot.xErr()
yErr = np.sqrt(y)
yErr[np.isclose(y, 0.)] = 0.
fig, ax = plt.subplots()
ax.errorbar(x, y, xerr=xErr, yerr=yErr, fmt="o", markersize=3.)
ax.set_ylabel("Counts/channel")
ax.set_xlabel("Channel")
ax.set_xscale("log")
ax.set_yscale("log")
plt.show()
\end{lstlisting}

Note that in PyXspec, the uncertainties change depending on the statistic being used. The plots that are produced are shown in Figure~\ref{fig:ex2}.

\end{boxI}

\begin{boxI}
\subsubsection*{Exercise 3 -- rebin the data}
To simplify the following exercises, for the purposes of this textbook Chapter, we will be fitting with WStat. In (Py)Xspec, WStat is activated by setting the fit statistic to CStat and loading a spectrum with its corresponding background. If there is no background model loaded, WStat is actually being used. 

To use WStat in a safe way, we have to bin the source spectrum so that the background will have some counts in each bin:\\

\textbf{\texttt{ftgrouppha} (in \texttt{ftools})}:
\begin{lstlisting}[frame=single]
ftgrouppha grouptype=bmin groupscale=5 clobber=yes \
infile=fpma_60ks.pha \
backfile=fpma_60ks_bkg.pha \
outfile=fpma_60ks_bmin5.pha
\end{lstlisting}

The remaining exercises use the created \texttt{fpma\_60ks\_bmin5.pha} spectral file  above.

\end{boxI}

\newpage
\section{Frequentist data analysis}\label{sec2}

\subsection{Fitting by minimization}

Now we can stick these statistics into some off-the-shelf minimizer.
These are general-purpose computer algorithms that begin from a user-provided
parameter guess, and iteratively walk around the parameter space
in such a way to minimize this ``cost'' function. Many minimizers
exist, the most common in X-ray spectral fitting being the
Levenberg-Marquardt algorithm, the Nelder-Mead simplex method and
differential evolution. Of these, Levenberg-Marquardt considers the
model gradient (which is however typically not available) to guess
the next point assuming a locally quadratic landscape. The assumption is optimal for the case of $\chi^{2}$ (note eq.~\ref{eq:chi2} is a quadratic polynomial), but not CStat. 
The simplex method evolves a group of
points, always replacing the worst fit with a new linear combination
of the remaining points, to wander towards the minimum, and is quite robust. 

In complicated models, there may be multiple but distinct regions
of the parameter space which provide comparable fit quality. Which
of these local optima is found by the minimizer is partially determined
by the initial guess and partially random. The properties above lead
to the unsatisfying activity of repeatedly restarting the minimizer
to another starting point, or strategies of scanning over one parameter
while optimizing the remainder to escape such local optima and discover
the global minimum in fit statistic. Fancier algorithms such as differential evolution
try to mitigate this problem by maintaining a large population
of points that searches the parameter space.

Whichever minimizing algorithm is employed, the algorithm iterates until a stopping criterion is reached. This is typically an estimate that any further refinements in the best-fit parameters only lead to improvements smaller than a threshold $\epsilon$. 

At the end of the minimization process, we have the parameters $\hat{\theta}$
which most frequently produce the observed data. For example, if we
fit a power law model $F=N\times(E/1\,\mathrm{keV})^{-\Gamma} \mathrm{erg/s/cm^{2}/keV}$
with two parameters, $\theta=(N, \Gamma)$, the minimizer may gleefully
report: $\Gamma=1.26$! As physicists, this is not the end of our
analysis because the next, and often more important,
question is: which $\Gamma$ values are ruled out by the data?

\begin{figure}
\begin{centering}
\includegraphics[width=0.99\textwidth]{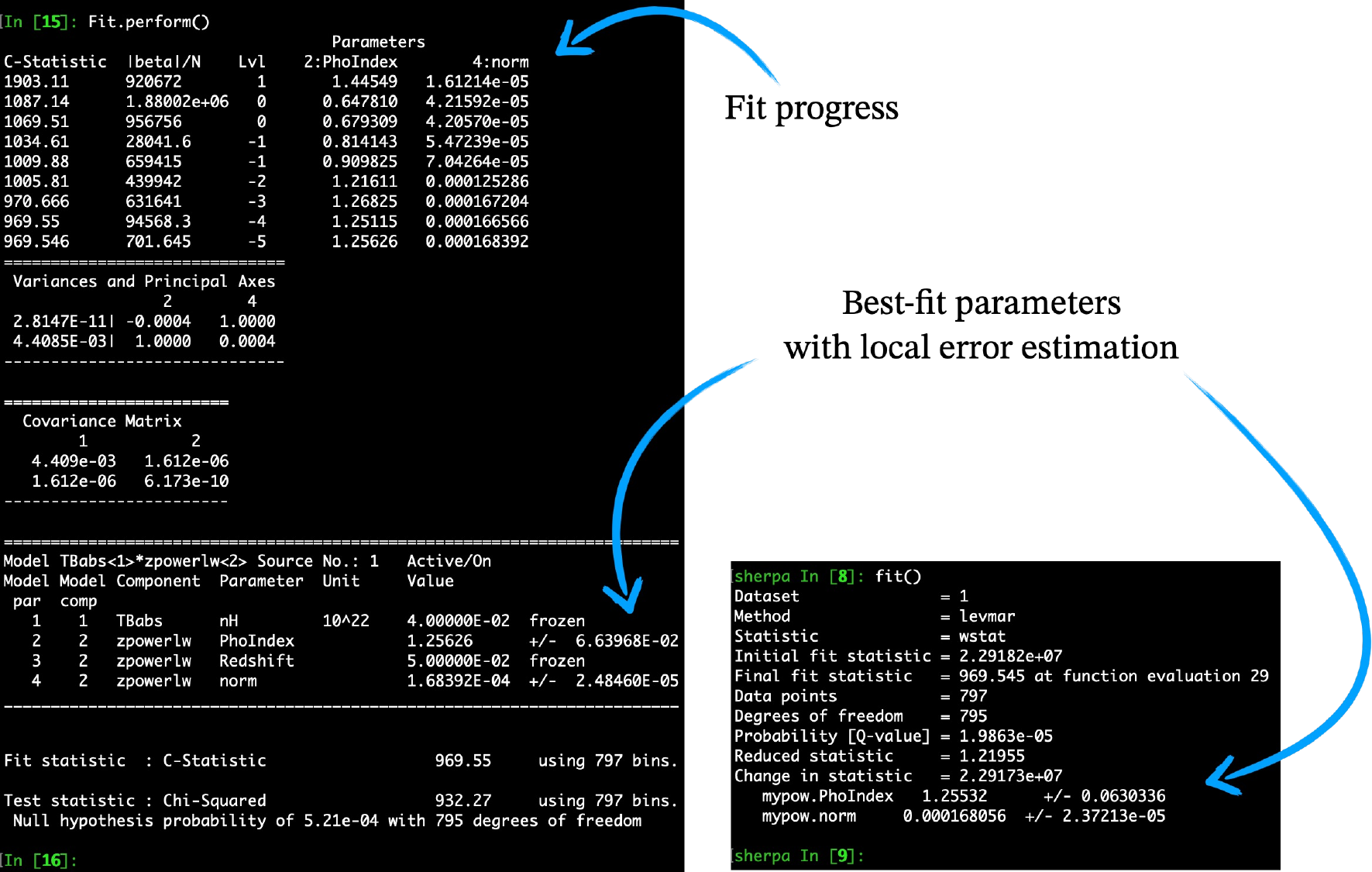}
\par\end{centering}
\caption{iPython output in PyXspec (left) and Sherpa (right) from the commands given in Exercise~4. A number of useful quantities are given in both, including the fit statistic value, number of degrees of freedom and corresponding parameter values. From the optimizer's limited exploration, also a rough uncertainty estimate around the (locally) best fit is given.\label{fig:ex4}}

\end{figure}

\begin{boxI}
\subsubsection*{Exercise 4 -- fitting a model to the data}
Now let's fit a model to the binned data created from the previous exercise. To start with, we will fit the data with a redshifted power law whilst also accounting for absorption from material along the line-of-sight in the Milky Way. Note the redshift of the source is 0.05 and the Galactic column density is $4\times10^{20}$\,cm$^{-2}$.\\

\textbf{Sherpa}:
\begin{lstlisting}[frame=single]
load_pha(1, "fpma_60ks_bmin1.pha")
ignore_id(1, "0.:3.,78.:")
set_analysis(1, "ener", "rate")
set_stat("wstat")
set_xsabund("wilm")
model1 = xstbabs.nhgal*xszpowerlw.mypow
set_par(nhgal.nh, val=0.04, frozen=True)
set_par(mypow.phoindex, val=1.8, min=-3., max=10., frozen=False)
set_par(mypow.redshift, val=0.05, frozen=True)
set_par(mypow.norm, val=1.e-5, min=1.e-10, max=1.e-1, frozen=False)
set_source(1, model1)
plot_fit(xlog=True, ylog=True)
fit()
\end{lstlisting}

\textbf{PyXSpec}:
\begin{lstlisting}[frame=single]
from xspec import *
Plot.xAxis = "keV"
Plot.device = "/xw"
Xset.abund = "wilm"
AllData("1:1 fpma_60ks_bmin1.pha")
AllData.ignore("1: 0.-3. 78.-**")
Fit.statMethod = "cstat"
Fit.query = "yes"
model1 = Model("TBabs*zpowerlw")
model1.TBabs.nH.values = (0.04, -1.)
model1.zpowerlw.PhoIndex.values = (1.8, 0.1, -3., -2., 9., 10.)
model1.zpowerlw.Redshift.values = (0.05, -1.)
model1.zpowerlw.norm.values = (1e-5, 0.01, 1e-10, 1e-10, 1e-1, 1e-1)
Plot("ldata")
AllModels.show()
Fit.perform()
print(Fit.nIterations)
Fit.show()
\end{lstlisting}

The iPython output from these commands are shown in Figure~\ref{fig:ex4}.

\end{boxI}

\subsection{Frequentist error analysis}

To estimate uncertainties on our model parameters, let's assume a few
things are true. First, let's assume the model is linear in its parameters.
Secondly, let's assume we are always far from boundaries of the parameter
space. Thirdly, let's assume we are in the high-data regime, i.e., that every spectral bin has many counts and all model parameters can be constrained well. In that
(optimistic) case, considered by Wilks' theorem \citep{wilks1938large}, one can make a second-order Taylor expansion of the
statistic, in other words, it falls off quadratically. This is akin
to a Gaussian distribution. With this picture in mind, it makes sense
to ask what the $1\sigma$-equivalent parameter range is. This can
be estimated by the quadratic approximation that the minimizer already
built internally and reports as errors (see Figure~\ref{fig:ex4}). 

However, in many realistic cases, none of the three assumptions are fulfilled in X-ray astronomy. The approximation can be poor, most frequently producing under-estimated errors, for example because of non-linear degeneracies. 
Instead, it is common to scan the parameter space by varying each parameter in turn, while simultaneously optimizing the other parameters. This is called "profiling the likelihood" and illustrated in Figure~\ref{fig:chi2}.
Where the statistic has worsened by $\Delta=1$, the $1\sigma$ equivalent
is reached, and this defines the $1\sigma$ confidence intervals. Equivalently $\Delta=2$ for $2\sigma$-equivalent confidence intervals.

Now, {\em what precisely are confidence intervals supposed to describe
and do they}? If we assume that the optimal parameters obtained, $\hat{\theta}$,
are the true parameters of the process out there in the Universe, and
we generate many thousands of spectra that could have been observed,
and for each of them the minimum was determined and a $1\sigma$ confidence
interval constructed as described above, then the true value $\hat{\theta}$
is contained in $68\%$ of these confidence intervals. This is the definition of confidence intervals.

Since in many cases either the model is not linear, or we are not
in the high-data regime or we are not away from the parameter boundaries,
or we do not necessarily trust our minimizer to be perfect, this is
not quite right. Therefore, we have to actually simulate a thousand
spectra, fit them, and look how the confidence intervals actually
behave. We could \textit{calibrate} $\Delta$ so that the constructed
confidence intervals contain the input value $68\%$ of the time.
Then, we have the desired property without needing strong assumptions.
This is a typical activity for observing proposals, where we want to convince a panel that with the obtained spectrum, no matter the random realisations, we can constrain some parameter of interest with high probability.

\begin{boxI}
\subsubsection*{Exercise 5 -- the power of a proposed experiment}

For an observing proposal, you suspect that the true model parameter has a certain value, e.g., $\Gamma=1.4$, but it may also be $\Gamma=2.0$. We want to convince an observing panel that with 10\,ks of exposure time, the two scenarios can be distinguished.

Generate 1000 spectra by Monte Carlo sampling with a Poisson random number generator. Do this for each of the two scenarios. For each spectrum, perform a spectral fit and determine confidence intervals.
How often is the "wrong" value outside the confidence interval? This corresponds to the power of the experiment to distinguish the two scenarios.

Statistical subtleties to watch out for: Where should the fit be started? Is it okay to start at the known true value? Did you rebin each generated spectrum?



\end{boxI}

\subsection{Model checking}


We perform the above exercise and find $\Gamma=1.26\pm0.1$. Beautiful!
Then our advisor asks us to make a plot of the spectrum and the model
fit. So we plot the observed counts $d_{i}$ against the model predicted counts $\lambda_{i}$. Our heart sinks: these
look nothing alike! Our $\Gamma$ constraint is not meaningful.

This process is formalised in model checking, which tests whether
the model could produce the observed data. Importantly, the model
is considered in isolation, without alternatives.

The first approach is to consider a null hypothesis significance test,
where either the model defined is true, or another (unspecified) model
is true. We consider the data that could have been stochastically
produced by our model, and look whether the data the telescope actually
recorded are an implausibly infrequent outlier. We could generate thousands of
Poisson counts from the model, and count what fraction of these realisations
lie above the observed counts $c_{i}$, which is the p-value. In the
case of a Poisson distribution, this can also be computed analytically:
\begin{align}
p=\sum_{c_{i}'=c_{i}}^{\infty}P_{\mathrm{Poisson}}(c_{i}',\lambda_{i})
\end{align}
With the conversion between p-value and $\sigma$ from the Gaussian
distribution (a common convention), we would call it a $3\sigma$
outlier if $p=0.1\%$. However, if the channel bins are fine, each
bin may have little information to judge a model in this way. To address
this, we could rebin channels together for this test (see rebinning above). One may
be tempted to consider the cumulative count distribution, $C_{i}=\sum_{j=1}^{i}c_{j}$,
with a Kolmogorov-Smirnov or Anderson-Darling test. However, since
we determined $\hat{\theta}$ from the same data $c_{i}$ by optimization,
its p-values are not valid (see~\url{https://asaip.psu.edu/Articles/beware-the-kolmogorov-smirnov-test/}).
\cite{Kaastra2017} analysed how the CStat may be used directly as a measure of the goodness of fit.

The second approach is to visualise the data and think about it. You
can try various visualisations, such as grouping channels together
to get a clearer impression of the deviations, residual plots, cumulative
plots of $C_{i}$, etc. This will give you ideas of what could be wrong,
and suggest alternative models to try. How to compare among multiple
plausible models is explained below. Visualisations (and thinking)
are highly recommended approaches. Not everything has to be a test.

For further reading on model checking and detection of components via reduced chi square, F-test and likelihood ratio tests, and their pitfalls, we refer the reader to \cite{Protassov2002,Andrae2010}.

\begin{figure}
\begin{centering}
\includegraphics[width=0.99\textwidth]{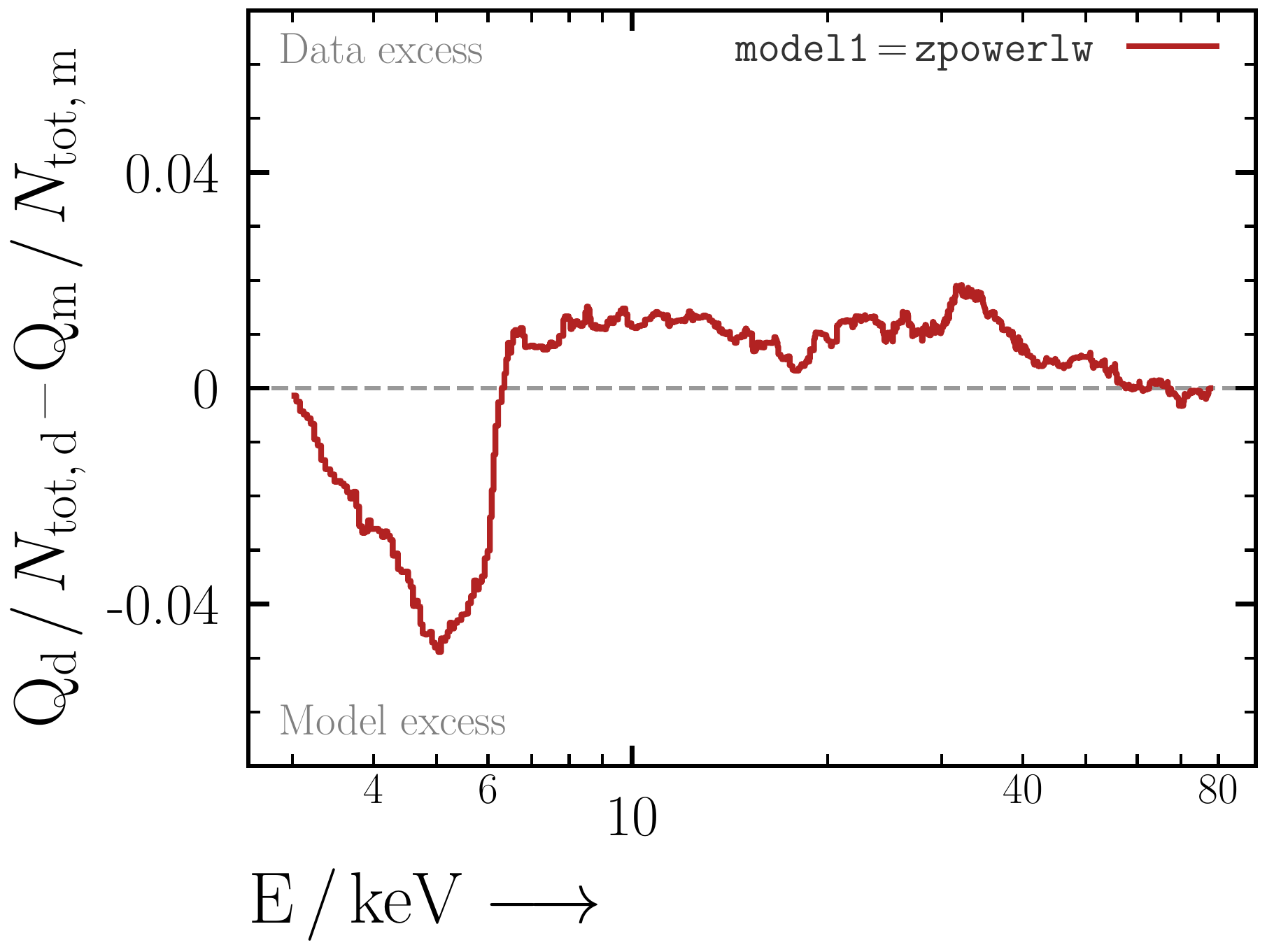}
\par\end{centering}
\caption{\label{fig:ex6} \lq Q\,--\,Q difference\rq\ plot for the frequentist Levenberg-Marquardt fit of \texttt{model1} from Exercise~6. Plotted on the y-axis is the difference between the cumulative \lq data\rq\ counts (i.e. the cumulative source\,$+$\,background counts) and the cumulative model (including the background) counts. The clearest feature in the plot is a large deficit of counts $\lesssim$\,6\,keV. The cumulative has been normalised to unity. Cumulative-based plots have the benefit of not being dependent on binning for visual clarity, apart from the binning required for the fit itself.}
\end{figure}

\begin{boxI}
\subsubsection*{Exercise 6 -- plotting the cumulative data ($Q_{\rm d}$) \& cumulative model ($Q_{\rm m}$) counts}
There are many useful ways to visually verify a model fit to data. One such plot is the \lq Q\,--\,Q\rq\ plot -- traditionally a plot of cumulative data (i.e. source\,$+$\,background counts) vs. cumulative model counts (i.e. observed model\,$+$\,background counts). A perfect fit would be a 1:1 relation in such a plot, but can be difficult to associate a particular energy range to any discrepancy from a perfect fit. Instead, we will produce a \lq Q\,--\,Q difference\rq\ plot to check if the model~1 fit from Exercise~4 makes sense. The only difference is that we will plot the energy vs. the difference of the two cumulative distributions (each separately normalised to 1). The code below leads on from Exercise~4 to save the source, background and model counts with \texttt{pandas}.\\

\textbf{Sherpa}:
\begin{lstlisting}[frame=single]
import pandas as pd
set_analysis(1, "ener", "counts")
df_data = {}
data = get_data_plot(1)
model_fit = get_fit_plot(1)
df_data["E_keV"] = data.x
df_data["E_keV_err"] = data.xerr
df_data["sb_counts"] = data.y
df_data["mb_counts"] = model_fit.y
df = pd.DataFrame(data = df_data)
df.loc[:, "Qsb"] = df["sb_counts"].cumsum() / df_data["sb_counts"].sum()
df.loc[:, "Qmb"] = df["mb_counts"].cumsum() / df_data["mb_counts"].sum()
df.to_csv("ex6_QQ.csv", index=False)
\end{lstlisting}

\textbf{PyXSpec}:
\begin{lstlisting}[frame=single]
import pandas as pd
import numpy as np
Plot.background = True
Plot("lcounts")
df_data = {}
df_data["E_keV"] = np.array(Plot.x())
df_data["E_keV_err"] = np.array(Plot.xErr())
df_data["s_counts"] = np.array(Plot.y())
df_data["b_counts"] = np.array(Plot.backgroundVals())
df_data["m_counts"] = np.array(Plot.model())
df_data["sb_counts"] = df_data["s_counts"] + df_data["b_counts"]
df_data["mb_counts"] = df_data["m_counts"] + df_data["b_counts"]
df = pd.DataFrame(data = df_data)
df.loc[:, "Qsb"] = df["sb_counts"].cumsum() / df_data["sb_counts"].sum()
df.loc[:, "Qmb"] = df["mb_counts"].cumsum() / df_data["mb_counts"].sum()
df.to_csv("ex6_QQ.csv", index=False)
\end{lstlisting}

Note the columns of each dataframe are subtly different -- Sherpa does not subtract the background from the source\,$+$\,background counts automatically, so we do not save the background separately.

The dataframe can then be used to plot \texttt{df["E\_keV"]} vs. \texttt{df["Qsb"]}\,--\,\texttt{df["Qmb"]} as shown in Figure~\ref{fig:ex6}.
\end{boxI}

\begin{boxI}
\subsubsection*{Exercise 7 -- Adding complexity with model~2}
Clearly from Exercise~6 and Figure~\ref{fig:ex6}, something is missing in model~1. Due to the large deficit of counts $\lesssim$\,6\,keV, an educated guess would be to include absorption. Thus below we first define and fit a new model~2:\\

\textbf{Sherpa}:
\begin{lstlisting}[frame=single]
model2 = xstbabs.nhgal*xszTBabs.zTBabs*xscabs.cabs*xszpowerlw.mypow
set_par(nhgal.nh, val=0.04, frozen=True)
set_par(zTBabs.nh, val=10., min=0.01, max=1000., frozen=False)
set_par(zTBabs.redshift, val=0.05, frozen=True)
set_par(cabs.nh, val=zTBabs.nh)
set_par(mypow.phoindex, val=1.8, min=-3., max=10., frozen=False)
set_par(mypow.redshift, val=0.05, frozen=True)
set_par(mypow.norm, val=1.e-5, min=1.e-10, max=1.e-1, frozen=False)
set_source(1, model2)
plot_fit(xlog=True, ylog=True)
fit()
\end{lstlisting}

\textbf{PyXSpec}:
\begin{lstlisting}[frame=single]
model2 = Model("TBabs*zTBabs*cabs*zpowerlw")
model2.TBabs.nH.values = (0.04, -1.)
model2.zTBabs.nH.values = (10., 0.1, 0.01, 0.01, 1000., 1000.)
model2.zTBabs.Redshift.values = (0.05, -1.)
model2.cabs.nH.link = "p2"
model2.zpowerlw.PhoIndex.values = (1.8, 0.1, -3., -2., 9., 10.)
model2.zpowerlw.Redshift.values = (0.05, -1.)
model2.zpowerlw.norm.values = (1.e-5, 0.01, 1.e-10, 1.e-10, 1.e-1, 1.e-1)
Plot("ldata")
AllModels.show()
Fit.perform()
\end{lstlisting}
\end{boxI}

\subsection{Model comparison}\label{sec:freqmodcomp}

\begin{figure}
\begin{centering}
\includegraphics[width=0.7\textwidth]{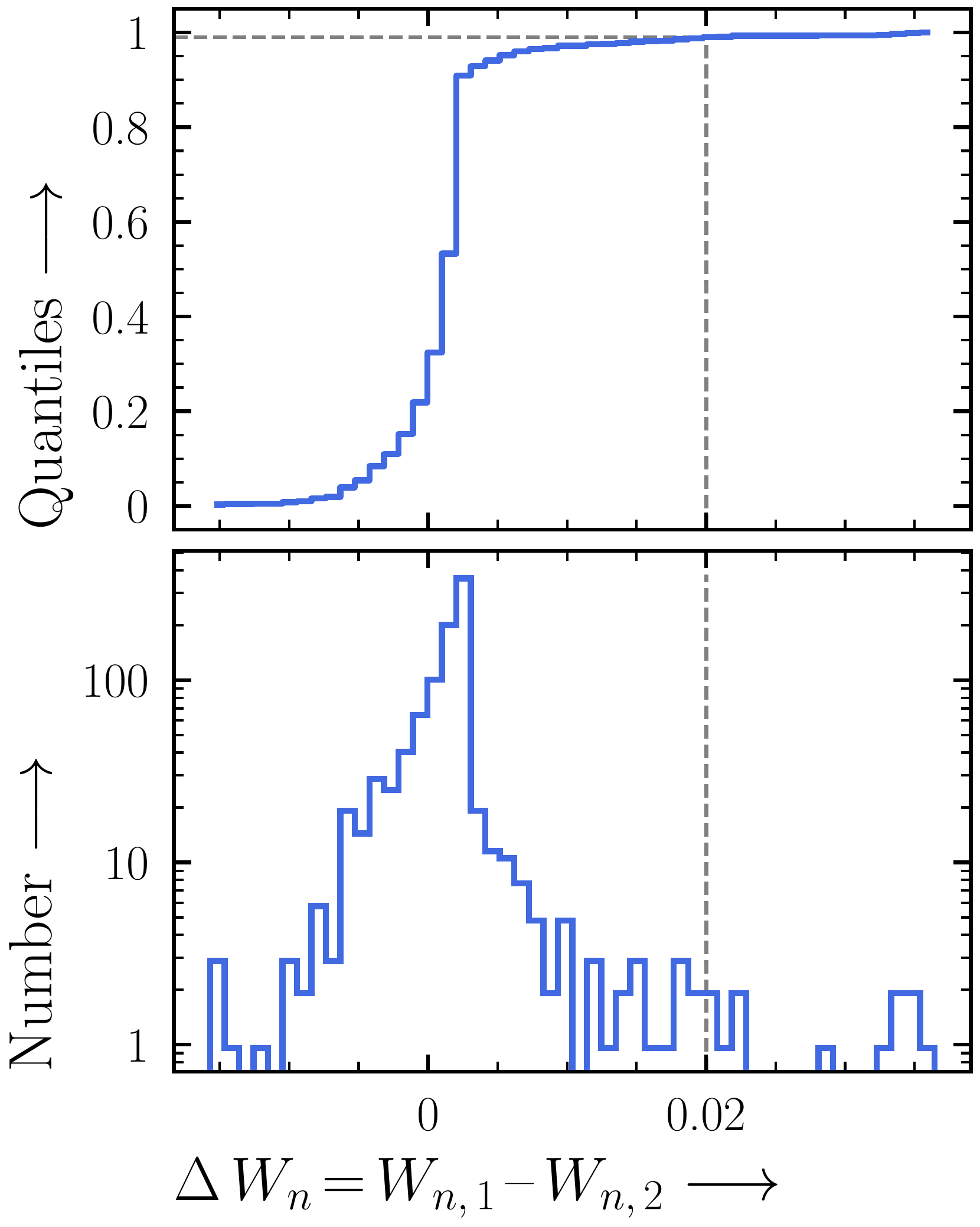}
\par\end{centering}
\caption{\label{fig:ex81} \textbf{False positive (type~I) error.} Following Exercise~8.1 (and Table~\ref{tab:confusionmatrix}) to estimate the false positive (i.e. type~I) error, we simulate from the simpler model~1 and re-fit with both model~1 and model~2. Here we plot the resulting distribution of $\Delta W_{n}$\,=\,$W_{\rm model\,2}\,/\,n_{\rm model\,2}$\,--\,$W_{\rm model\,1}\,/\,n_{\rm model\,1}$ from the simulations. The 99th quantile from the distribution is a value of 0.02 -- thus if we choose this value of $\Delta W_{n}$ for our threshold, we would expect to select the more complex model~2 $\sim$\,1\% of the time in the event that the null model~1 was correct.}
\end{figure}

Finally, we can consider the situation where two models of physical
processes of equal a-priori plausibility are to be judged by the data.
Is model A significantly better than model B? This is also a case
of null hypothesis significance testing, if model A is the null hypothesis,
and we want to reject model A in favour of the alternative hypothesis B.
This has two possible outcomes: either the rejection is successful,
in which case model B is picked because there was enough evidence
for model B. Or the rejection is unsuccessful, in which case model
A is kept, either because it is better or because there was not enough
distinguishing evidence. Again, the null hypothesis significance test
needs a confidence threshold, e.g., $p<1\%$. This sets how often
we would erroneously prefer model B over model A even when model A
was true (false positive rate, or type I error, see Table~\ref{tab:confusionmatrix} and Figure~\ref{fig:ex81}). 

\begin{figure}
\begin{centering}
\includegraphics[width=0.7\textwidth]{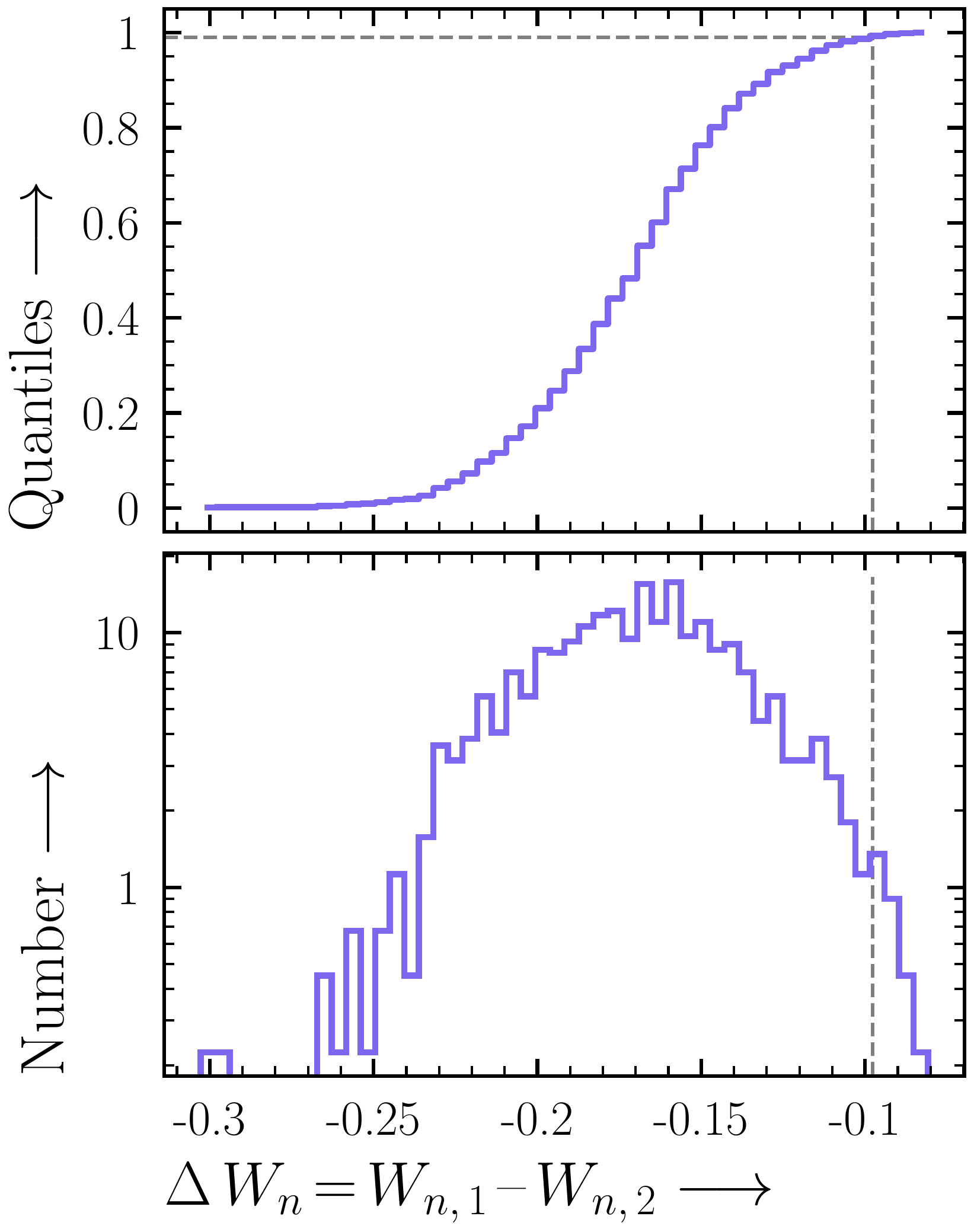}
\par\end{centering}
\caption{\label{fig:ex82} \textbf{False-negative (type~II) error.} In contrast to Exercise~8.1, we now simulate from the more complex model~2 before re-fitting with both model~1 and model~2. Following the same prescription as in Figure~\ref{fig:ex81}, we plot the resulting distribution of $\Delta W_{n}$, and find the 99th quantile to be 0.098. Thus if we choose this to be our threshold, we would expect to preferentially select model~1 $\sim$\,1\% of the time, when the more complex model~2 was correct. We can also say that at this $\Delta W_{n}$ threshold, the power is 99\%.}
\end{figure}

In contrast, the false negative
rate (type II error) is the erroneous preference of model A when
model B was true. Ideally, one would like both errors to be small,
but this depends on the model test (see Figure~\ref{fig:ex82}).

\begin{table}
    \centering
    \begin{tabular}{l|l l}
         \hfill \textbf{Output $\rightarrow$}  & Model A found & Model B found \\         \textbf{True input $\downarrow$} \hfill & (negative) & (positive) \\
         \hline
         \hline
         Simpler model A \statslanguage{(null)} & true negative & false positive \statslanguage{(type I error, $\alpha$)} \\
         Complex model B \statslanguage{(alternative)} & false negative \statslanguage{(type II error, $\beta$)} & true positive \statslanguage{(power)} \\
    \end{tabular}
    \caption{Confusion matrix. Even if the simpler model A were always true (first row), the decision procedure under a certain threshold may falsely claim a detection of model B, giving a false positive. Alternatively, even if model B is actually true (second row), the data may not provide enough evidence to the decision procedure to prefer model B over model A, giving a false negative. These concepts apply to any method for making decisions, but have different names in different disciplines. The terms used by statisticians are in \statslanguage{blue}.}
    \label{tab:confusionmatrix}
\end{table}

In the idealised situation considered above (linear model, high-data
regime, away from the parameter bounds), which does not hold in X-ray
astronomy, there are analytic statistical tests which consider the
statistic difference ($\Delta\mathrm{CStat}=\mathrm{CStat}_{A}-\mathrm{CStat}_{B}$
or $\chi_{A}^{2}-\chi_{B}^{2}$) between the two models. These include
the F-test and likelihood ratio test. Do not use these. Do not trust
the results of these, because their assumptions are not valid for
our data.

You can however build a reliable null hypothesis test from what we
have already discussed: simulate 1,000 spectra from the null model
(model A), fit each with both models, and compute the statistic difference
$\Delta\mathrm{CStat}$. This gives you a distribution of $\Delta\mathrm{CStat}$
values expected if the null model was true. In rare cases, by chance,
model B may be more prone to produce the observed data, $\Delta\mathrm{CStat}>0$. Reading off the $99\%$ quantile, we obtain
a $\Delta\mathrm{CStat}$ threshold $\Delta\mathrm{CStat}_{\mathrm{crit}}$,
which sets the type I error to $1\%$. This is a subtle point, but
follows from the definitions: we have simulated under model A, and
in 1\% of cases this $\Delta\mathrm{CStat}_{\mathrm{crit}}$ is exceeded.
In other words, if we apply our calibrated criterion $\Delta\mathrm{CStat}>\Delta\mathrm{CStat}_{\mathrm{crit}}$,
and select model B, the purity of this process is 99\%.

To obtain the power of our test to distinguish the models (1 - false
negative rate), we would have to simulate under model B, and count
the fraction of cases where we indeed select model B.

\begin{boxI}
\subsubsection*{Exercise 8.1 -- false positive rate (type~I error)}
Now that we have defined and fit model~1 as well as the (more complex) model~2, we can calculate the false positive rate, i.e. how many times we would expect model~2 to be chosen when the simpler model~1 was actually correct. Simulate 1,000 spectra from model~1 and fit each with both model~1 and model~2. Then record the $\Delta\mathrm{WStat}$ each time to build a distribution that is expected if the null model (model~1) were true.\\

\end{boxI}

\begin{boxI}
\subsubsection*{Exercise 8.2 -- false negative rate (type~II error)}
From Exercise~8.1, we know the fraction of times we would expect to select model~2 when the simpler model~1 was actually correct. Next, calculate the opposite -- how many times do we expect to select model~1 when the more complex model~2 was actually correct?
\end{boxI}

\subsection{Limitations so far}\label{sec:frequentistlimitations}

As you have seen above, analytic derivations make unreasonable approximations.
By generating many simulated spectra and analysing them, it is however
possible to get answers and be confident in our inference
for excluding parameter values, and distinguishing models.

The frequentist approach describes the reliability of procedures to
recover input values. In the process however, the best-fit value $\hat{\theta}$
was assumed to be the true value. What we would rather have, indeed,
is a procedure which tells us what the true value out there in the
Universe is, or at least a probability distribution over the true
value. For example, we would like to make statements -- starkly different
to those possible with confidence intervals (re-read above) -- such
as: 
\begin{itemize}
\item the true value of $\Gamma$ lies between 2.1 and 2.2 with 99\% probability
\item the true value of $\Gamma$ lies above 2.5 with 0.1\% probability.
\end{itemize}
In other words, we are making statements of the integrated probability
over some interval of a parameter, $\int_{a}^{b}P(\theta)\,d\theta$.
By the way, when you set $a=b$ the integral becomes zero, because
the probability to have exactly that value is infinitesimally small
compared to all possible values; but we can live with that. 

Probability distribution functions (PDFs) allow interesting ways to
judge parameter ranges relative to each other. Unfortunately, there is no
inference procedure that just produce these. The key reason is that to define $\int P(\theta)\,d\theta$
we assumed that we are integrating with equal weight over $\theta$.
If we changed our definition of our power law model to be $N\times E^{-\log\gamma}$
instead of $N\times E^{-\Gamma}$, we would obtain the same optimal
parameter $\log\gamma=\Gamma$, but $\int P(\Gamma)\,d\Gamma\neq\int P(\gamma)\,d\gamma=\int P(\Gamma)\,d\log\Gamma$.

Therefore, we need one more assumption to define probability distributions. Namely, we need to specify how ``large" a region of the parameter space is relative to another
region. This is known as the integration measure. In the context
of the following it is also called a prior (ooooohhhh scary!). Beware
if someone tells you they can, without assuming a prior make PDFs
or statements like the above. There are assumptions hidden that define
their prior, and they are unaware of what their inference procedure
is actually doing. 

Now we can introduce the mathematics of producing PDFs, namely Bayesian inference. Further below, we will combine the strengths of frequentist and Bayesian paradigms.

\begin{boxI}
\subsubsection*{Exercise 8.3 -- model~3}
The next exercises will feature an additional model~3, which includes a Gaussian emission line with the same redshift as the source. The syntax is as follows.

\textbf{Sherpa}:
\begin{lstlisting}[frame=single]
model3 = xstbabs.nhgal*(xszTBabs.zTBabs*xscabs.cabs*xszpowerlw.mypow
                        +xszgauss.line)
set_par(nhgal.nh, val=0.04, frozen=True)
set_par(zTBabs.nh, val=10., min=0.01, max=1000., frozen=False)
set_par(zTBabs.redshift, val=0.05, frozen=True)
set_par(cabs.nh, val=zTBabs.nh)
set_par(mypow.phoindex, val=1.8, min=-3., max=10., frozen=False)
set_par(mypow.redshift, val=0.05, frozen=True)
set_par(mypow.norm, val=1.e-5, min=1.e-10, max=1.e-1, frozen=False)
set_par(line.linee, val=6.4, min=6., max=7.2, frozen=False)
set_par(line.sigma, val=1.e-3, min=1.e-3, max=1., frozen=False)
set_par(line.redshift, val=0.05, frozen=True)
set_par(line.norm, val=1.e-5, min=1.e-10, max=1.e-1, frozen=False)
set_source(1, model3)
plot_fit(xlog=True, ylog=True)
fit()
\end{lstlisting}

\textbf{PyXSpec}:
\begin{lstlisting}[frame=single]
model3 = Model("TBabs*(zTBabs*cabs*zpowerlw+zgauss)")
model3.TBabs.nH.values = (0.04, -1.)
model3.zTBabs.nH.values = (10., 0.1, 0.01, 0.01, 1000., 1000.)
model3.zTBabs.Redshift.values = (0.05, -1.)
model3.cabs.nH.link = "p2"
model3.zpowerlw.PhoIndex.values = (1.8, 0.1, -3., -2., 9., 10.)
model3.zpowerlw.Redshift.values = (0.05, -1.)
model3.zpowerlw.norm.values = (1.e-5, 0.01, 1.e-10, 1.e-10, 1.e-1, 1.e-1)
model3.zgauss.LineE.values = (6.4, 0.1, 6., 6., 7.2, 7.2)
model3.zgauss.Sigma.values = (1.e-3, 0.01, 1.e-3, 1.e-3, 1., 1.)
model3.zgauss.Redshift.values = (0.05, -1)
model3.zgauss.norm.values = (1.e-5, 0.01, 1.e-10, 1.e-10, 1.e-1, 1.e-1)
Plot("ldata")
AllModels.show()
Fit.perform()
\end{lstlisting}

Note we choose the minimum and maximum line energy to encompass the neutral Fe\,K$\alpha$ fluorescence line.
\end{boxI}

\newpage
\section{Bayesian inference}\label{sec3}

\subsection{Terminology}

Bayes' theorem uses the likelihood function to update an auxiliary
probability density, the prior PDF $P(\theta)$, to obtain a posterior
PDF $P(\theta|D)$:
\begin{equation}
P(\theta|D)=\frac{P(D|\theta)P(\theta)}{P(D)}\label{eq:bayestheorem}
\end{equation}
This is a reordering of the terms in the law of conditional probability,
with specific meanings assigned. The first term in the numerator is
the likelihood, which specifies the frequency of producing a data
set $D$ given assumed parameters $\theta$, ${\cal L}(\theta)=P(D|\theta)$
(specifically interpreted as \lq the probability of D given $\theta$\rq). The second term is the prior.
Let's call it $\pi$, so that not everything is called $P$, then we have:
\begin{align}
P(\theta|D)=\frac{{\cal L}(\theta)\times\pi(\theta)}{Z}
\end{align}
The denominator normalises the posterior, $Z=\int{\cal L}(\theta)\times\pi(\theta)d\theta$,
and is known as the Bayesian evidence. It is also known as the marginal
likelihood, because the process of integrating away a variable $P(a)=\int P(a,b)\,da$
is called marginalisation.

\subsection{Parameter estimation}

\begin{figure}
\begin{centering}
\includegraphics[width=0.99\textwidth]{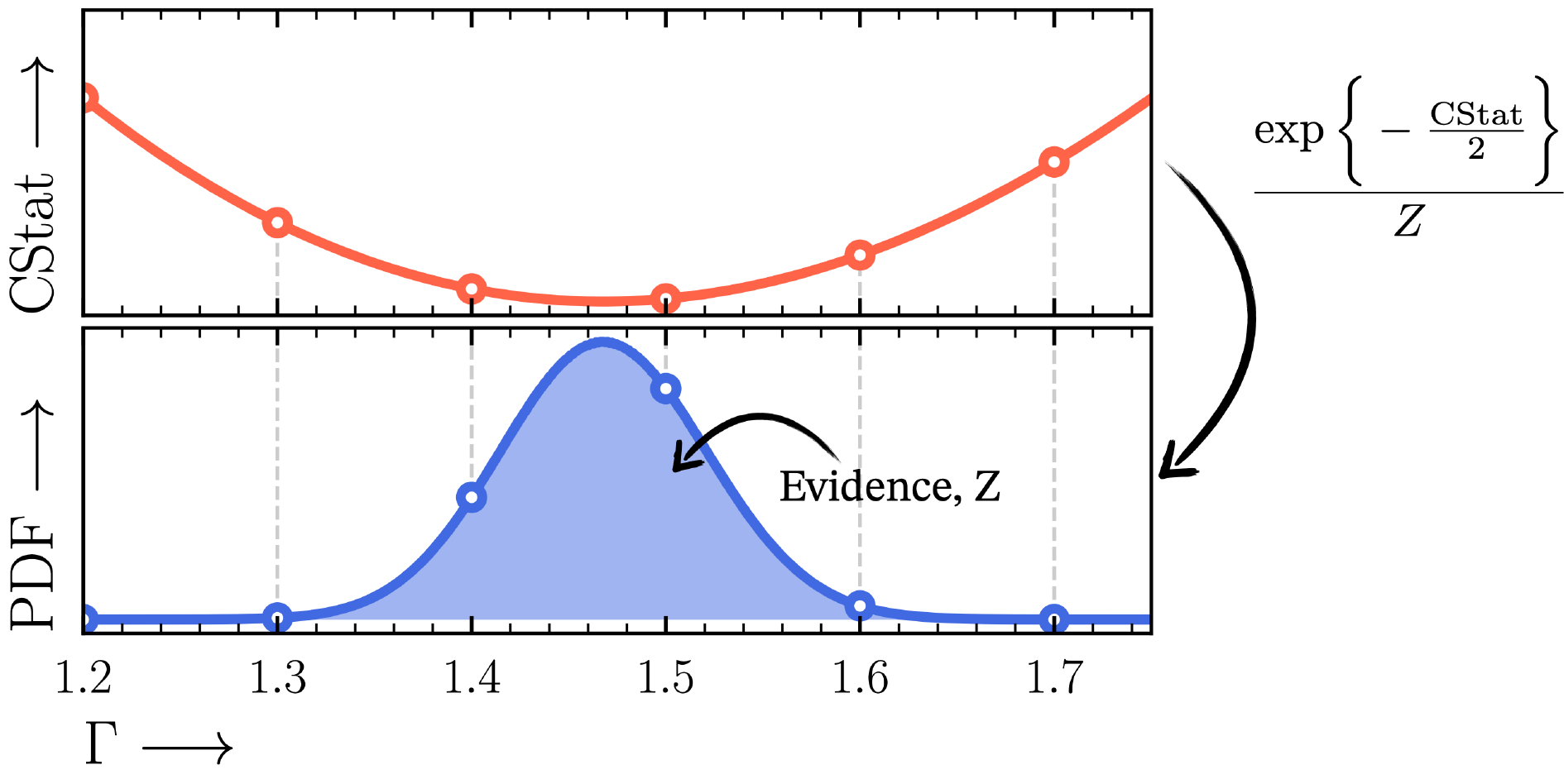}
\par\end{centering}
\caption{\label{fig:bayesgrid}\textbf{Bayesian inference on a grid.} Stepping through
the model parameter $\Gamma$, we obtain a CStat value at each grid
point (red curve in the upper panel). Converting CStat to a probability ($\exp\left\{ -\mathrm{CStat}/2\right\} /Z$)
and normalising gives the posterior probability distribution (blue curve in the lower panel).
The normalising constant is the Bayesian evidence, $Z$.}

\end{figure}


So if we make a grid in $\Gamma$, compute ${\cal L}(\Gamma)\times\pi(\Gamma)$
at each grid point and numerically integrate it, are we doing Bayesian
parameter estimation? Yes. Almost everything else is numerical details.
As shown in Figure~\ref{fig:bayesgrid}, from a grid we can
compute $Z$ with numerical integration, and then we can identify
the grid intervals that contain 99\% of the probability, or ask how
much probability is above $\Gamma>2.5$. If we also had the normalisation
as a second parameter, we would first marginalise it away and get the marginal
posterior distribution of the $\Gamma$ parameter:
\begin{align}
P(\Gamma|D)=\frac{\int\pi(\Gamma)\pi(N){\cal L}(\Gamma,N)\,dN}{\int\pi(\Gamma)\pi(N){\cal L}(\Gamma,N)\,dN\,d\Gamma}
\end{align}

\subsection{Choosing priors}

Specifying priors is a well-studied problem, and there are several
approaches which can also be combined. One approach is to pick an
informative prior from previous knowledge, for example from simulations,
plausibility arguments or from previous studies. For example, we know
today from surveys of Active Galactic Nuclei (AGN) in the local Universe,
that the photon index describing the approximate intrinsic shape of the X-ray coronal spectrum is $\sim$1.8--2 with some scatter (e.g., \citealt{Ricci17_bassV}). Thus a reasonable prior for X-ray spectral fitting of an AGN intrinsic (i.e. absorption-corrected) spectrum could be Gaussian distributed with mean 2 and 0.2 standard
deviation.

Another approach is to define uninformative priors. Here, one considers
all possible outcomes one can conceive before the experiment, and
assigns them probabilities based on the principle of least surprise,
the principle of maximum entropy, or such that rescaling does not
change under reparameterization of the model. Skipping many technical
details here, two special cases are common, which derive uniform priors
in linear (for parameters with unknown offset, such as the location
of an unknown Gaussian emission line) or logarithmic coordinates (such
as the same Gaussian line's dispersion and normalisation, or more generally other parameters of unknown
scale).

\begin{figure}
\begin{centering}
\includegraphics[width=0.7\textwidth]{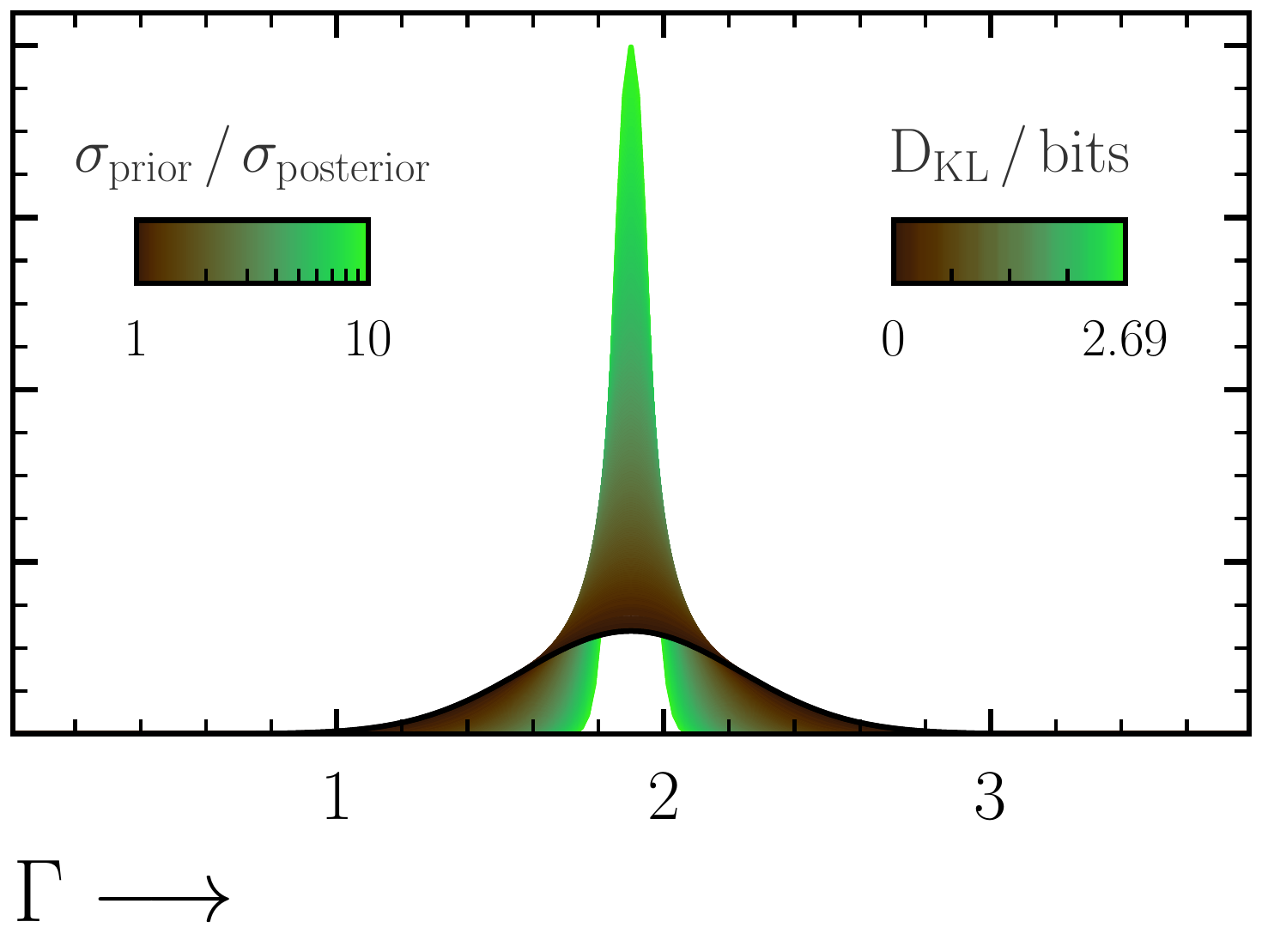}
\par\end{centering}
\caption{\label{fig:infogain}\textbf{Measuring information gain with the Kullback-Leibler divergence.} A Gaussian prior on some parameter, $\Gamma$ is assumed with fixed width, and the information gain is calculated for a series of posteriors with narrower widths than the original prior. For reference, 1\,bit of information gain corresponds to a Gaussian shrinking its standard deviation by a factor of three \citep{Buchner22_infogain}.}
\end{figure}

It is good practice to try several priors that the author or readers 
may consider reasonable and test how sensitive the results of the study
are to the assumption.
In general, the posterior PDF depends on the prior, but in some situations
it really does not. This is the case when the likelihood function
is highly concentrated and forces the probability to go there. In
this case, we say that the data are highly informative. Then, trying
different priors gives the same result, as long as the prior has support where needed, i.e., the probability is not zero. 
How informative the data is is formalised with the \lq information gain\rq\ or Kullback-Leibler divergence. It estimates how much the posterior changed (usually shrunk) compared to the prior distribution, typically measured in bits. For details and how to interpret the resulting numbers, see \citealt{Buchner22_infogain} and Figure~\ref{fig:infogain}.

\subsection{Computation in multiple dimensions}
With the necessary concepts of Bayesian inference, probability peaks in parameter spaces and information gain introduced, we now discuss computational issues that occur when models have many parameters. The number of model parameters, $d$, is the dimensionality of the space over which the PDF and its integration is defined. With increased parameter dimensions, uniform grids become exponentially
costly to evaluate. The exponential behaviour arising from testing all parameter combinations is known as the curse of dimensionality. Smarter algorithms are needed. The most common
of these is Markov Chain Monte Carlo (MCMC), and more recently
Nested Sampling Monte Carlo. Here ``Monte Carlo" refers to the use of random numbers to compute results. Both techniques estimate the posterior
distribution, and produce several thousands of equally probable posterior
samples, $\theta_{1},\theta_{2},...,\theta_{i}$, which can then be
used by summation to perform the integrals mentioned above. Nested
sampling additionally computes $Z$, which is useful for Bayesian
model comparison (see Section~\ref{sec:ns}).

\subsubsection{Markov Chain Monte Carlo}
Let's first consider MCMC. We begin with a starting point $\theta_{i}$
with $i=1$, and an auxiliary proposal function, for example a narrow Gaussian
distribution centred at our starting point. We draw a proposed point
$\theta_{i}'$ from the proposal function, and compare the posterior ratio of the two: $\alpha=\frac{{\cal L}(\theta')\times\pi(\theta')}{{\cal L}(\theta)\times\pi(\theta)}$
\citep{Metropolis1953}. If the posterior is higher at the proposed
point $\theta'$, i.e., $\alpha>1$, we set $\theta_{i+1}=\theta'_{i}$.
If it is lower, we still jump there $\theta_{i+1}=\theta'_{i}$ with
probability $\alpha$, and otherwise remain at the current point $\theta_{i+1}=\theta_{i}$.
This Metropolis procedure creates a chain of points $\theta_{i}$
whose distribution converges, given infinite iterations, to the (unknown)
posterior distribution.

Since computing time is finite, convergence is not guaranteed.
There are several diagnostics that can be used to test whether a limited
chain or several chains are problematic. Here we recommend in particular
the $\hat{R}$ diagnostic \citep{gelman1992single,Vehtari2019}, which
compares the scatter observed within chunks of the chain, to the scatter
observed across multiple, independently run chains. Additionally,
visual inspection of the parameters with iteration (i.e. a trace plot) can indicate chains
that are stuck and thus not usable. For a modern environment for the
analysis of MCMC, we recommend \texttt{arviz}\footnote{\url{https://www.arviz.org/}}.


\begin{figure}
\begin{centering}
\includegraphics[width=0.99\textwidth]{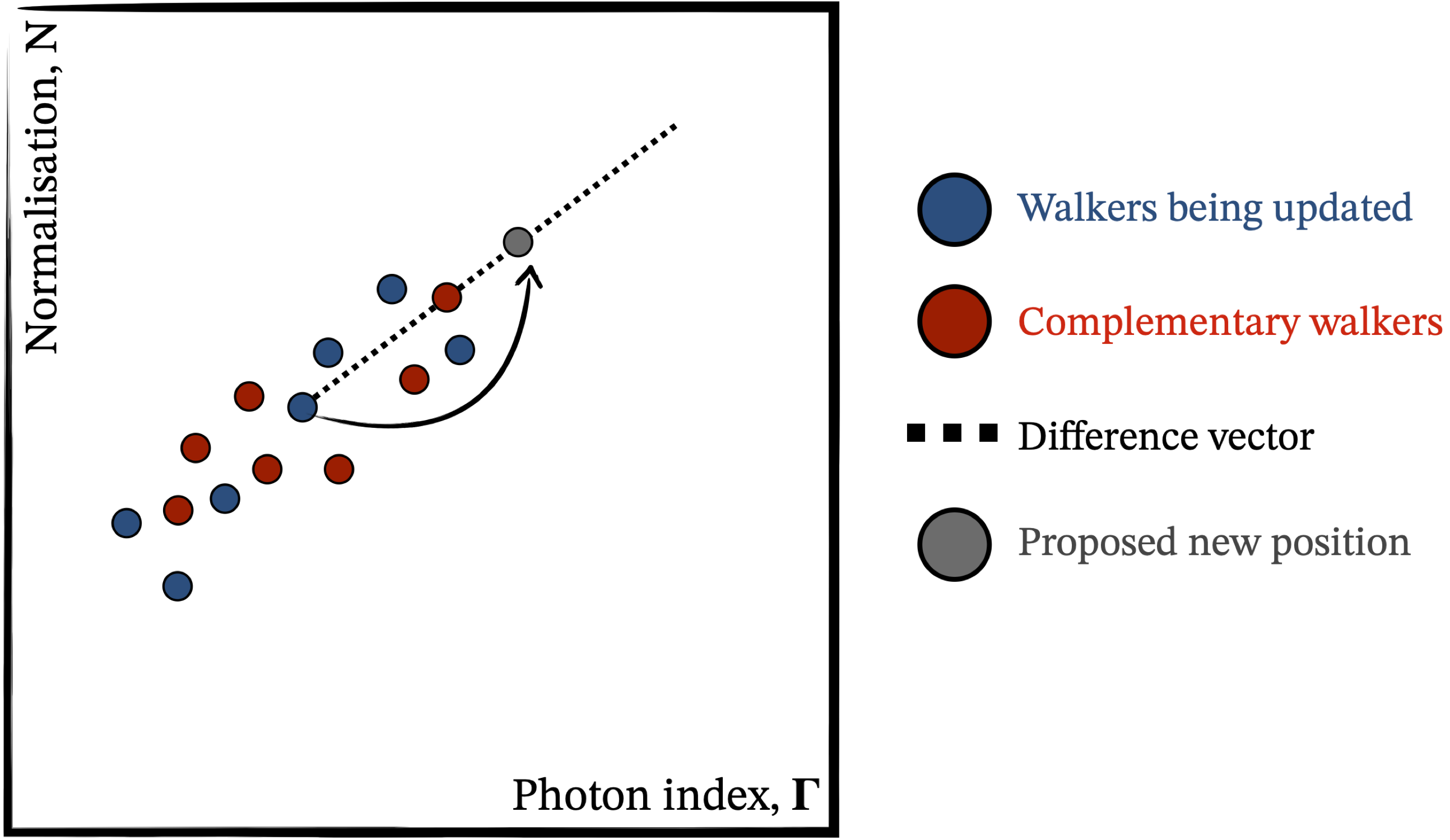}
\par\end{centering}
\caption{\label{fig:aies}\textbf{MCMC proposal of the affine-invariant ensemble sampler.}
Walkers are divided into two groups (blue and red). The blue walker paired with a red walker defines a difference vector (grey dashed), along
which a new position (grey point) for the walker is proposed. The proposal
is therefore oriented and scaled similar to how the walkers are already
distributed. Affine transformations of the parameter space preserve the proposal.}
\end{figure}

How efficient an MCMC method is depends strongly on the proposal for
the next point, or more generally the transition kernel. State-of-the-art techniques use Hamiltonian dynamics
to move with high acceptance rates through the space. However, gradients
of the models with respect to the parameters are necessary to achieve
this, which are commonly not available in current packages. Instead,
gradient-free MCMC algorithms include the popular Affine-Invariant
Ensemble Sampler (AIES) from \citet{2010CAMCS...5...65G}. Here, instead
of a single point performing a guided random walk, a population of
$K$ walkers is maintained. Half of the walkers are updated to a new
random position, using information from the other half (see Figure~\ref{fig:aies}). In particular,
the difference vectors to a randomly paired walker is used to propose
in that direction, with a random scale factor distributed around the
AIES scale factor parameter $\gamma$. The proposal is accepted or
rejected as described above. AIES proposes along lines, and the prior
is encoded in the parameterization along which lines are drawn, or
in the acceptance rule. In the case of AIES, each walker performs
a guided random walk, and the random walks are interdependent because
the walkers use each other as a proposal. Therefore, to diagnose the
chains produced, multiple independent runs need to be performed and compared, to
diagnose the runs with $\hat{R}$. See \url{https://johannesbuchner.github.io/autoemcee/mcmc-ensemble-convergence.html} for more technical details.

AIES can be well-behaved, and you
should expect a few 100,000 model evaluations or more (especially for complex models) to convergence,
which can be prohibitive for very slow models. AIES and in general
all MCMC techniques can get stuck in local optima. For more details on MCMC and its use in X-ray spectral analysis, see \cite{Dyk2001}.


\begin{figure}
\begin{centering}
\includegraphics[width=0.99\textwidth]{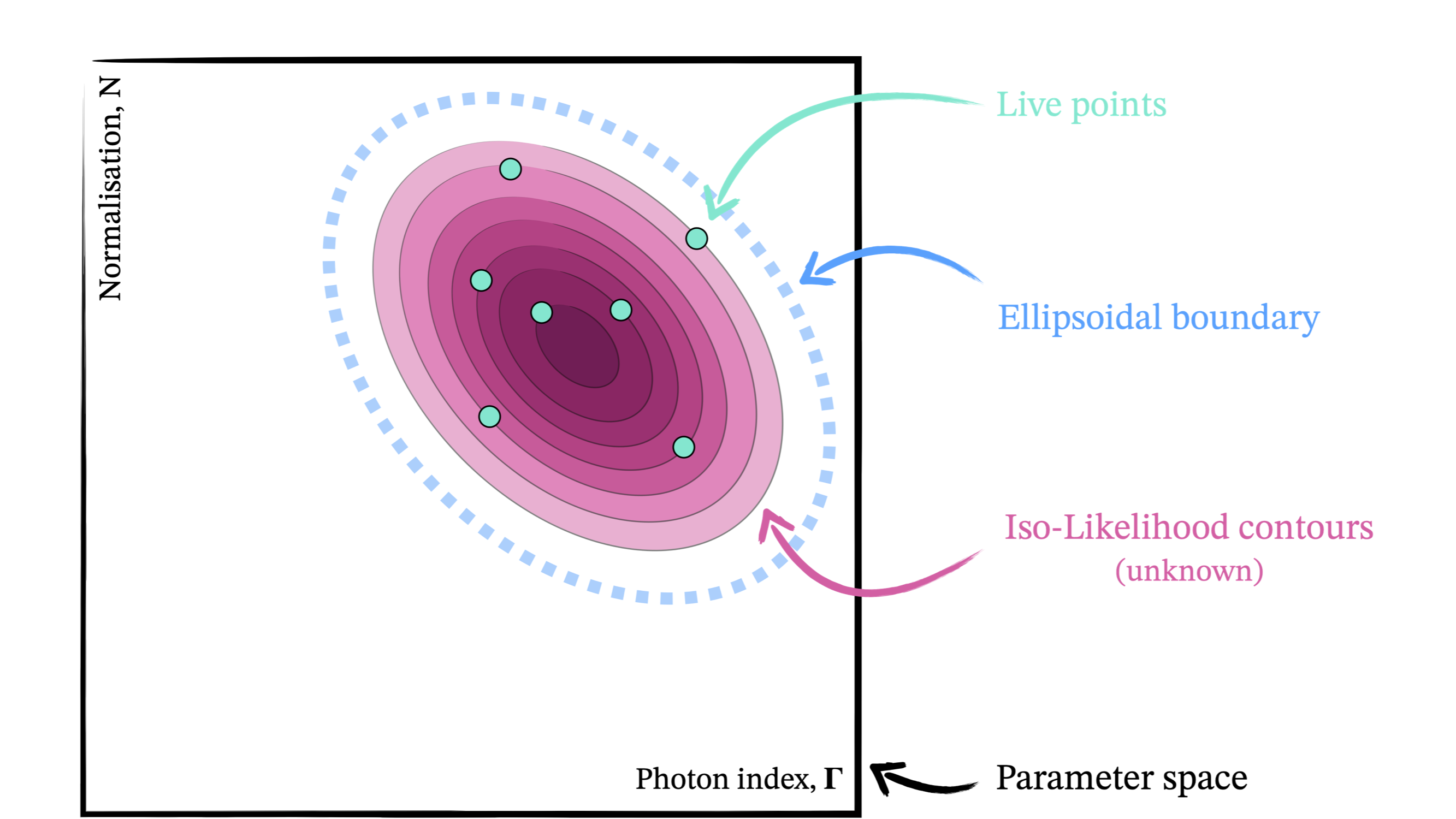}
\par\end{centering}
\caption{\label{fig:nsellipsoid}\textbf{Ellipsoidal Nested Sampling method.} The lowest likelihood live point is removed and a new
sample is to be drawn, with the constraint to lie above the outer (dashed)
live point contour line. However, this contour line is in general not known.
Drawing from the entire prior space (black rectangle), and rejecting
unsuitable points would be inefficient. Instead, an enlarged ellipsoid
(blue solid line) around the existing live points (cyan circles) is constructed. Then
points are proposed within the ellipsoid and accepted if above the
likelihood threshold. This works very efficiently when the contours are approximately ellipsoidal.}
\end{figure}

\subsubsection{Nested Sampling}\label{sec:ns}
Nested sampling Monte Carlo \citep{Skilling2004,Ashton2022} tackles
the harder task of integrating the entire prior-defined parameter space. Initially,
$K$ live points are sampled from the prior PDF, and are thus widely
distributed across the parameter space. Most of these give very poor
fits, with low likelihood ${\cal L}$. The worst (lowest likelihood) live point
with ${\cal L}_{1}$ is then discarded, and a new live point sampled
from the prior. However, only prior samples with ${\cal L}>{\cal L}_{1}$
are accepted as a replacement. This requirement causes approximately $1/K$ of the prior volume to be excluded. The procedure of discarding points and resampling them is repeated (nested sampling). At each iteration, the contribution to the evidence
integral is estimated as $w_{i}={\cal L}_{i}V_{i}$, i.e., the likelihood
weighted with the volume removed at iteration $i$, $V_{i}=\left(1-\frac{1}{K}\right)^{i}\frac{1}{K}$.
The evidence is $Z_{i}\approx\sum_{i}w_{i}$, and the
contribution of the live points is $Z_{\mathrm{live}}=\sum_{i}{\cal L}_{i}V_{i}$.
Since each nested sampling iteration removes a poor fit and allows
only better ones, after many iterations only very good fits remain.
These are concentrated in a comparatively tiny volume, and have ultimately similar
likelihoods, so that since $V\rightarrow0$, $Z_{\mathrm{live}}\rightarrow0$
and therefore $Z_{i}$ stabilizes and the iterating can be stopped.
Posterior samples of equal weight are produced by randomly sampling
the discarded points proportional to $w_{i}$. 

The difficult task 
of drawing a new live point under the restriction that the likelihood
must improve can be solved \citep[see][for a review]{Buchner2021a} by sampling in the neighbourhood of the
live points. Figure~\ref{fig:nsellipsoid} illustrates placing one
ellipsoid around them and sampling from the ellipsoid, while rejecting
proposals below the likelihood threshold \citep{Mukherjee2006}. Clustering
into multiple ellipsoids can be even more efficient \citep{Shaw2007,Feroz2008,Buchner2016}.
When the number of model parameters becomes large, MCMC can be
employed for this task inside nested sampling \citep{Skilling2004,Handley2015}, known as step samplers\footnote{See \url{https://johannesbuchner.github.io/UltraNest/example-sine-highd.html\#Step-samplers-in-UltraNest}.}. Both region sampling and step samplers are available through the Bayesian X-ray Analysis (BXA\footnote{\url{https://github.com/JohannesBuchner/BXA}}) package \citep{Buchner22_stepsamplers}.
For nested sampling, you should also expect a few 100,000 model evaluations
or more until completion.

\begin{figure}
\begin{centering}
\includegraphics[width=0.99\textwidth]{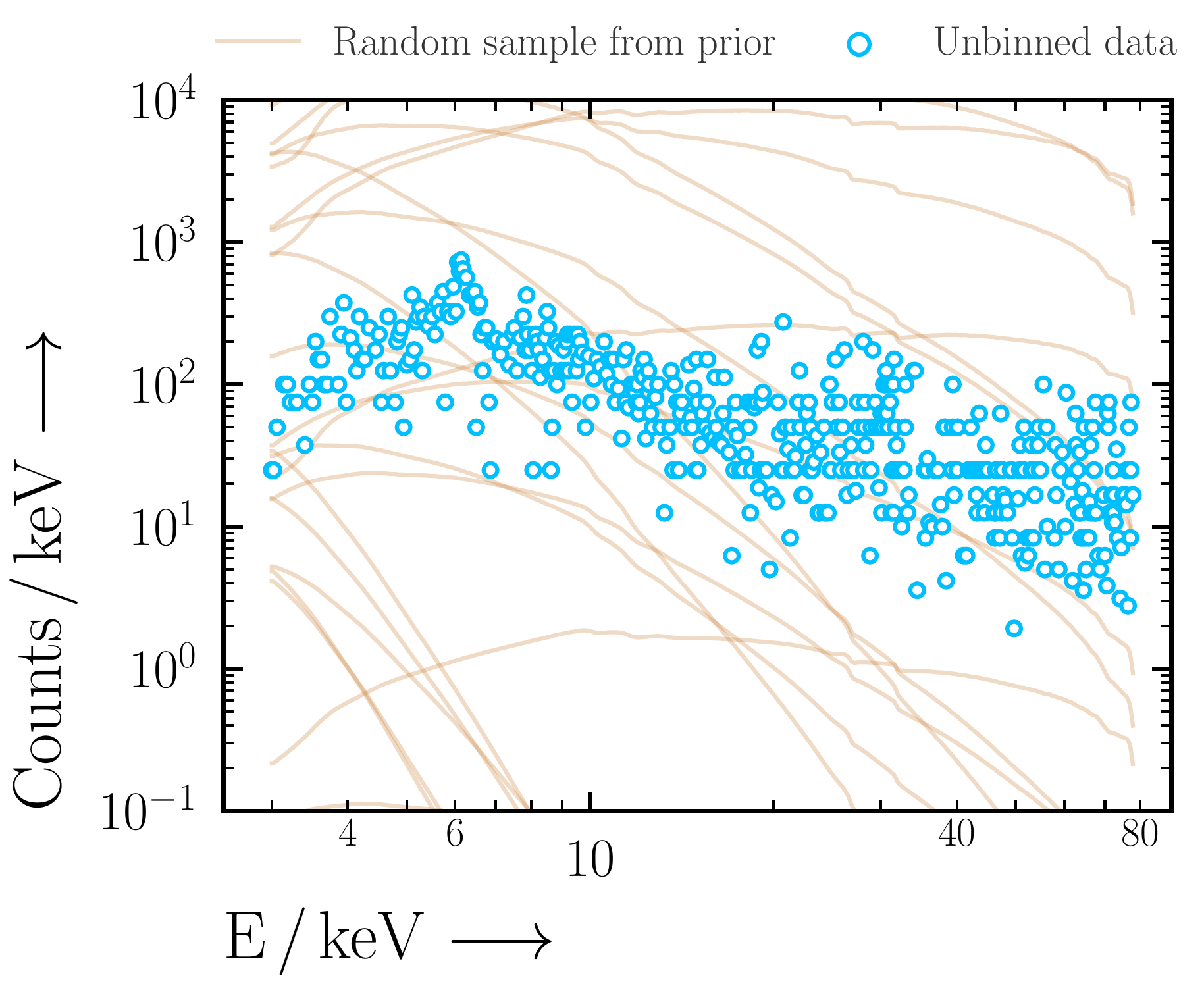}
\par\end{centering}
\caption{\label{fig:ex101}\textbf{Prior predictive checks.} Prior predictive checks are used to ensure the prior makes physical sense with regard to the data. By randomly sampling from the prior, we can see that the model is physically reasonable and is somewhat capable of producing the approximate shape of the data. The blue data points show the observed data.}
\end{figure}

\begin{boxI}
\subsubsection*{Exercise 9.1 -- prior predictive checks}
Now we will use nested sampling in our X-ray fits with BXA \citep{Buchner2014}, which connects the UltraNest \citep{ultranest} nested sampling algorithm to Sherpa \& PyXspec. Our first step after defining a model (as described in Exercises~4,~7~\&~8.3) is to define the priors for each parameter. To help us with this decision, we will use prior predictive checks to visually check that our priors make sense for the data we have.\\

\textbf{Sherpa} commands:
\begin{lstlisting}[frame=single]
import bxa.sherpa as bxa
import numpy as np
prior1 = bxa.create_uniform_prior_for(model1, model1.zpowerlw.PhoIndex)
prior2 = bxa.create_loguniform_prior_for(model1, model1.zpowerlw.norm)
solver = bxa.BXASolver(transformations=[prior1, prior2],
                       outputfiles_basename="model1_pyxspec")
for i in range(100):
    values = solver.prior_transform(np.random.uniform(
                                    size=len(solver.paramnames)))
    for i, p in enumerate(solver.parameters):
        p.val = values[i]
    plot_fit(xlog=True, ylog=True)
\end{lstlisting}

\textbf{PyXSpec} commands:
\begin{lstlisting}[frame=single]
import bxa.xspec as bxa
from bxa.xspec.solver import set_parameters
import numpy as np
prior1 = bxa.create_uniform_prior_for(model1, model1.zpowerlw.PhoIndex)
prior2 = bxa.create_loguniform_prior_for(model1, model1.zpowerlw.norm)
solver = bxa.BXASolver(prior=bxa.create_prior_function([prior1, prior2]),
                       parameters=[param1, param2],
                       outputfiles_basename="model1_sherpa")
for i in range(100):
    values = solver.prior_function(np.random.uniform(
                                   size=len(solver.paramnames)))
    set_parameters(transformations=solver.transformations, values=values)
    Plot("lcounts")
\end{lstlisting}
Save the output model spectral counts and overplot with the real counts to verify that the prior makes physical sense (see Figure~\ref{fig:ex101}). Try this for the remaining models~2~\&~3.
\end{boxI}

\begin{boxI}
\subsubsection*{Exercise 9.2 -- fitting a model to a spectrum with BXA}
After we have verified our priors with prior predictive checks in Exercise~9.1, we will now use BXA to fit model~1 to the spectrum.\\

\textbf{Sherpa} commands:
\begin{lstlisting}[frame=single]
results_model1 = solver.run(resume=True)
\end{lstlisting}

\textbf{PyXSpec} commands:
\begin{lstlisting}[frame=single]
results_model1 = solver.run(resume=True)
\end{lstlisting}

\texttt{results\_model1} contains the posterior samples (\texttt{results\_model1["samples"]}) for each parameter and the Bayesian evidence (\texttt{results\_model1["logZ"]}, \texttt{results\_model1["logZerr"]}), which should be printed to the screen after the fit is completed. You can also interact directly with the posterior samples yourself with e.g., \texttt{pandas}:

\begin{lstlisting}[frame=single]
import pandas as pd
df = pd.DataFrame(data=results_model1["samples"],
                  columns=solver.paramnames)
df.describe()
\end{lstlisting}

Next, try fitting each of the three models to the data.
\end{boxI}

\begin{boxI}
\subsubsection*{Exercise 10 -- more advanced usage with BXA}
For model 3, the BXA fit can take some time to converge. In general, for more complex fits (e.g., multiple datasets simultaneously, many parameters, high signal-to-noise spectra, etc.), the run time can be decreased with:\\

\textbf{Parallelisation:}\\
To parallelise BXA over \texttt{N} cores:
\begin{lstlisting}[frame=single]
mpiexec -np N python3 myscript.py
\end{lstlisting}

\textbf{Changing the arguments for \texttt{solver.run()}:}\\
To manually set the cut-off condition for the integration (high values will terminate quicker but will be less effective at finding multiple modes):
\begin{lstlisting}[frame=single]
frac_remain=0.5
\end{lstlisting}

To set the number of times \texttt{run()} will try to assess how many more samples are required:
\begin{lstlisting}[frame=single]
max_num_improvement_loops=0
\end{lstlisting}

To use a Reactive Nested Sampler (recommended):
\begin{lstlisting}[frame=single]
speed="safe"
\end{lstlisting}

To use a Step sampler with adaptive steps:
\begin{lstlisting}[frame=single]
speed="auto"
\end{lstlisting}

To use a step sampler with an integer number of steps (note this is faster but one should ensure the evidence estimation is accurate by performing multiple fits with differing numbers of steps):
\begin{lstlisting}[frame=single]
speed=int
\end{lstlisting}

Now try using a different sampling algorithm with a BXA \texttt{solver}. First, acquire the log-Likelihood, parameter names and priors:\\

\textbf{PyXSpec} commands:
\begin{lstlisting}[frame=single]
loglike = solver.log_likelihood
paramnames = solver.paramnames
prior = solver.prior_function
\end{lstlisting}

\textbf{Sherpa} commands:
\begin{lstlisting}[frame=single]
loglike = solver.log_likelihood
paramnames = solver.paramnames
prior = solver.prior_transform
\end{lstlisting}

Then pass these to other sampling algorithms (in addition to BXA Nested Sampling), such as MCMC:
\begin{lstlisting}[frame=single]
from autoemcee import ReactiveAffineInvariantSampler
sampler = ReactiveAffineInvariantSampler(paramnames, loglike, prior)
result = sampler.run()
\end{lstlisting}
\end{boxI}

\subsection{Using posteriors}


\begin{figure}
\begin{centering}
\includegraphics[width=0.99\textwidth]{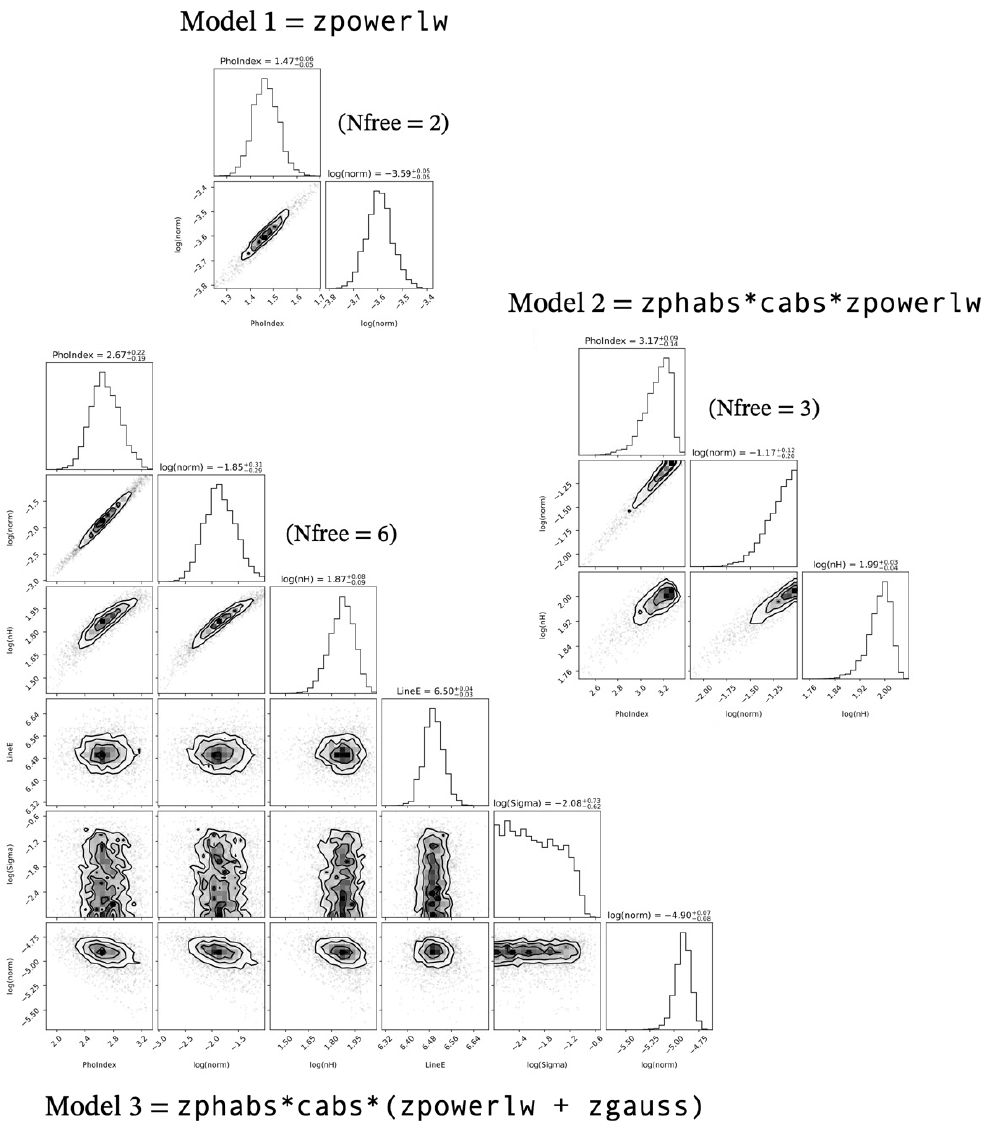}
\par\end{centering}
\caption{\label{fig:corner}\textbf{Corner plots created from the 3 models considered in this textbook chapter.} The first two models use a redshifted power law -- the first with free photon index and the second with additionally free normalisation. The second model considers an additional line-of-sight absorption component and the third considers an further redshifted Gaussian line. In all corner plots, marginal posterior probability distributions
are in the diagonal and pair-wise conditional posterior distributions (for models with $>$\,1 free parameter)
on the bottom left. The titles show median and $1\sigma$ equivalent
quantiles of the distribution. The contours enclose
12, 39, 68 and 86 per cent of the posterior probability (see \url{https://corner.readthedocs.io/en/latest/pages/sigmas/}).}
\end{figure}

Exercise 9.2 illustrates how to obtain posteriors for the model parameters. Figure~\ref{fig:corner} shows corner plots, which include one-dimensional histograms of the posterior samples (in the diagonal of the corner plot), and two-dimensional histograms, also known as marginal posterior distributions and conditional posterior distributions, respectively. 
Each parameter constraint is summarized, using the quantiles corresponding to the median and 1-sigma equivalent of the marginal posterior distribution (i.e. 50, 16 and 84 percentiles). 
The fraction of posterior samples above or below a threshold can be used to probabilistically answer interesting questions about where the true (unknown)
value lies, and where it does not (see~\ref{sec:frequentistlimitations}). 

\begin{figure}
\begin{centering}
\includegraphics[width=0.78\columnwidth]{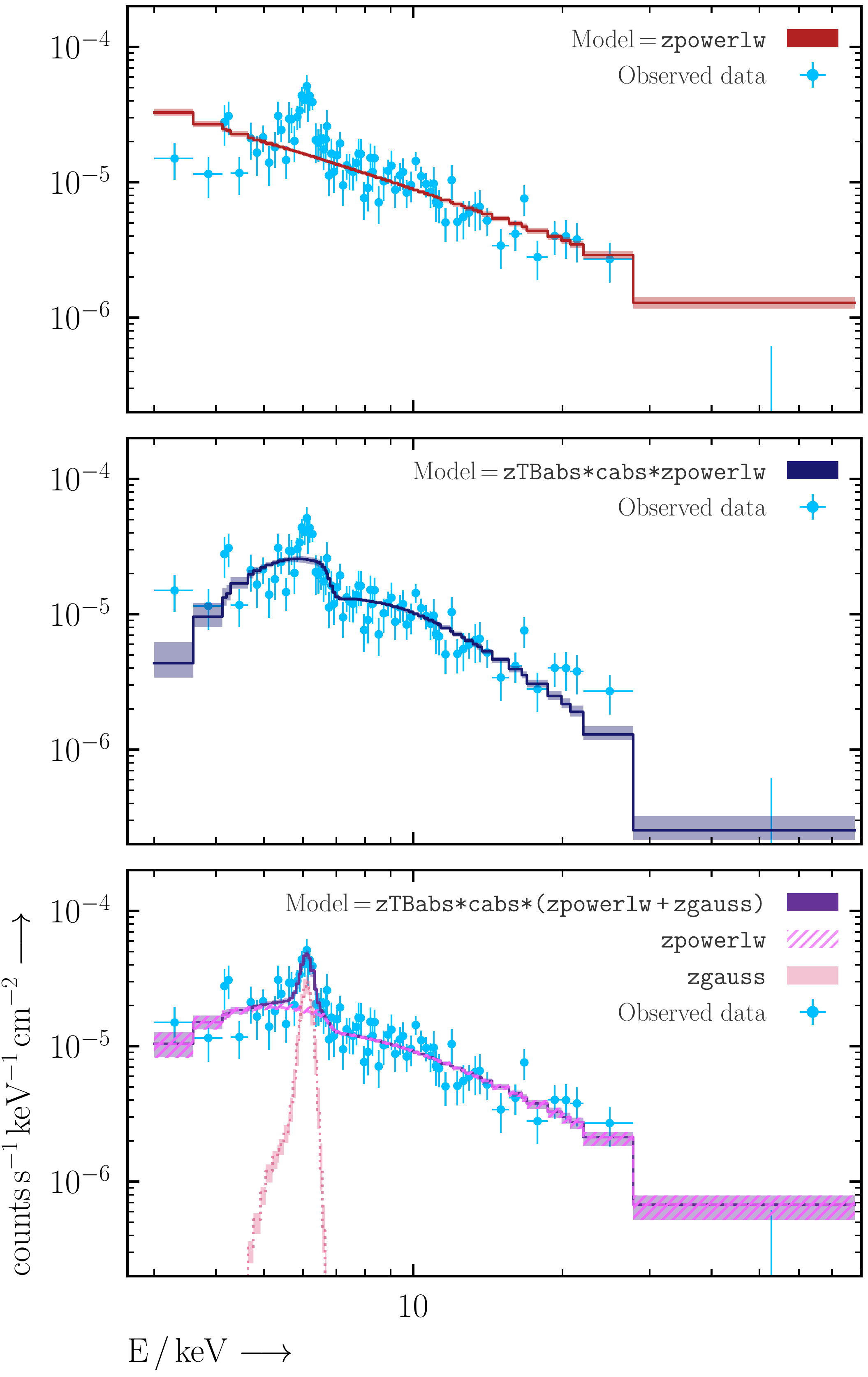}
\par\end{centering}
\caption{\label{fig:ldata_area} \textbf{A plot of the folded model for each of model~1,~2 and~3.} The posterior model range derived with BXA is shown as a shaded region on the model, together with the data points in each channel. The x-axis shows the nominal channel energy, and the y-values have been multiplied to correct for the bin width (keV), exposure time (s) and sensitive detector area (cm$^{-2}$). The last step lets the powerlaw source model (top panel) appear as an approximately straight line.}
\end{figure}

\begin{figure}
\begin{centering}
\includegraphics[width=0.78\columnwidth]{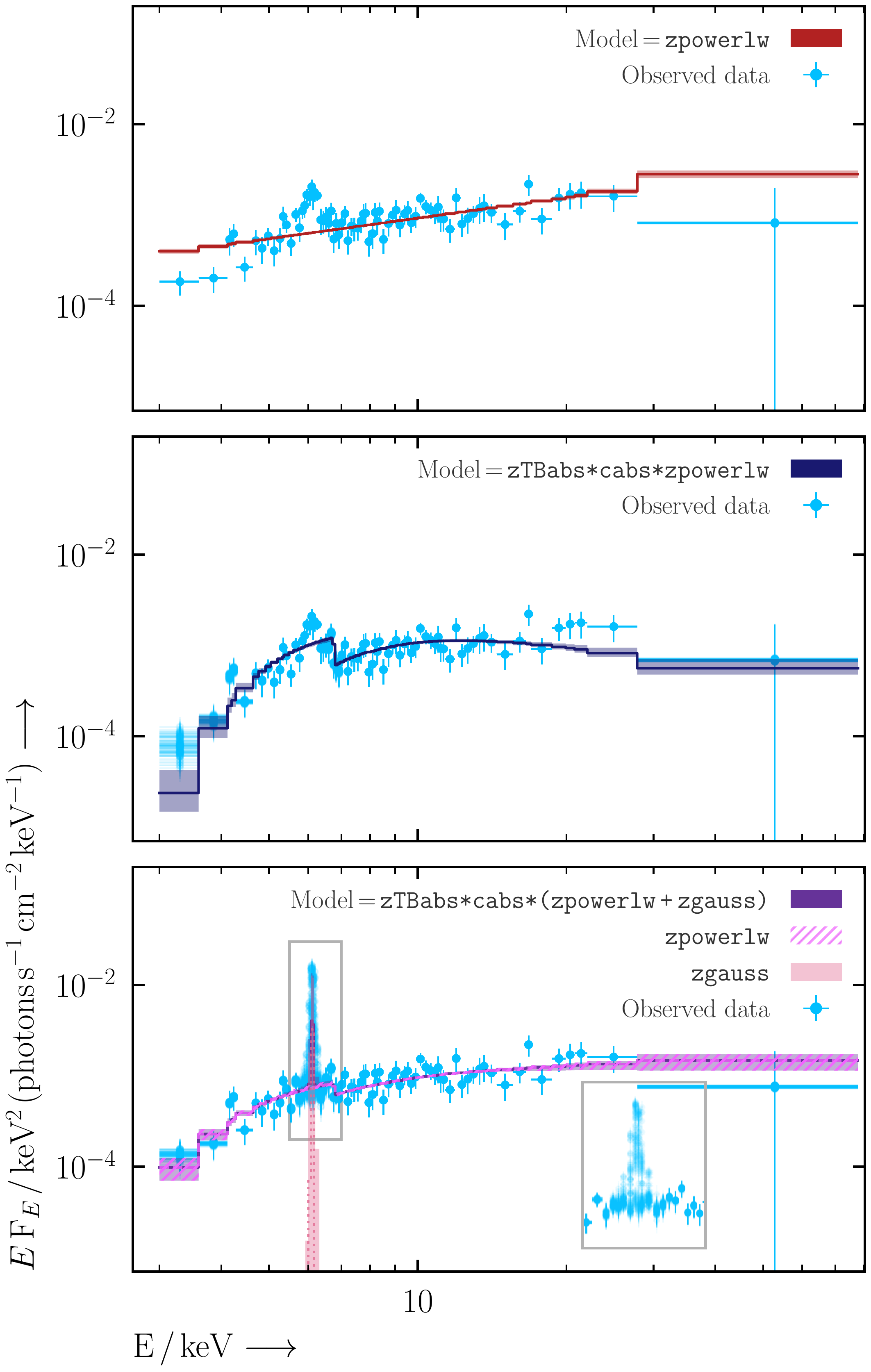}
\par\end{centering}
\caption{\label{fig:eeuf_plot} \textbf{Flux density plots with model realisations drawn from the posteriors samples.} The plotted "data points" here depends on the data and the model, so here we show each posterior realisation of the \lq data\rq\ in the plot. Introducing the narrow Gaussian feature in the third panel modifies the location of the data points. This highlights the possible issues in interpreting component strengths in unfolded units.}
\end{figure}

But posteriors can do more than this. 
With the posterior $P(\theta|D)$, you can obtain uncertainties on
any function that depends on $\theta$. Exercise~11 illustrates this by obtaining uncertainties on a line equivalent width from Gaussian line and powerlaw continuum parameter posteriors.
For example, we can get a
probabilistic prediction on the source model 
by computing:
\begin{align}
F(E)=\int F(E|\theta)\times P(\theta|D)\,d\theta
\end{align}
This provides a view of the emission process occurring without folding through the instrument response $F(E|\theta)$ (the `unfolded' model).
With this, we can predict uncertainties on a flux in another energy band, perhaps
observed by a future experiment. In practical terms, we take one posterior sample, compute something and obtain a prediction. Taking the ensemble of predictions over a large number of randomly-sampled posterior samples can be summarized with uncertainties, as illustrated in Figures~\ref{fig:ldata_area} and \ref{fig:eeuf_plot}.

\begin{boxI}
\subsubsection*{Exercise 11 -- propagating posteriors into secondary parameters of interest}
In Exercise~9.2 we saw how to save the posteriors into separate pandas dataframes. Now it is straightforward to calculate posteriors of tertiary parameters that depend on the already-acquired posterior parameters by analytical functions. For example, in model~3 we have an emission line (\texttt{zgauss}) and continuum (\texttt{zpowerlw}). The equivalent width of an emission line is defined as the flux in the line (in the line normalisation in the zgauss parameterisation used here), divided by the value of the continuum at the line energy ($N\times E_\mathrm{line}^{-\Gamma}$). Thus we can create a posterior on equivalent width as follows:\\

\textbf{Sherpa} commands:
\begin{lstlisting}[frame=single]
df = pd.DataFrame(data=results_model3["samples"],
                  columns=solver.paramnames)
df.loc[:, "EW"] = (10**df["line.lognorm"] / 10**df["mypow.lognorm"])
                   * df["line.linee"] ** df["mypow.phoindex"]
\end{lstlisting}

\textbf{PyXSpec} commands:
\begin{lstlisting}[frame=single]
df = pd.DataFrame(data=results_model3["samples"],
                  columns=solver.paramnames)
df.loc[:, "EW"] = (df["zgauss.norm"] / df["zpowerlw.norm"])
                   * df["zgauss.LineE"] ** df["zpowerlw.PhoIndex"]
\end{lstlisting}
What are the 16th, 50th and 84th quantiles on equivalent width in this source?
\end{boxI}

\subsection{Model checking}

Similar to the model checking section above, here we can take each
$\theta$ with its associated probability and predict model counts
$c_{i}'$. Comparing the predicted model counts to the observed model
counts, using for example $3\sigma$ intervals, is called posterior
predictive checks (PPC).

\begin{figure}
\begin{centering}
\includegraphics[width=0.99\textwidth]{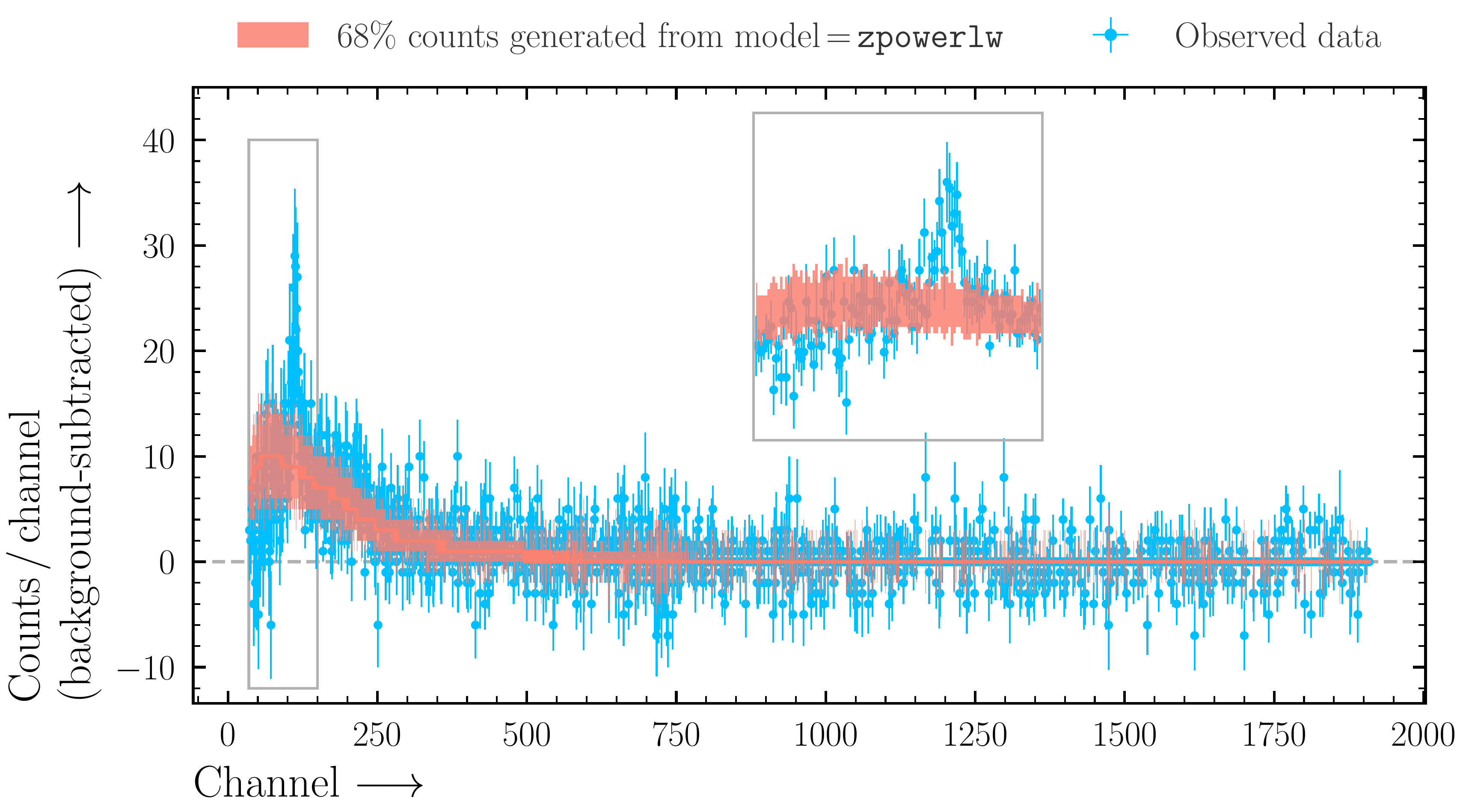}
\par\end{centering}
\caption{\label{fig:ppc_pow}\textbf{Model discovery with posterior predictive checks (PPC).} The observed data is shown in linear counts vs. channel with red errorbars. A number of posterior rows were sampled many times from a simple power law fit and data was simulated using the same conditions as the observed data (i.e. the same exposure, response, background file). The result is shown with a red filled region. As seen in the inset, at low channel values, there appears to be a deficit and surplus in counts at channels $\sim$\,25\,--\,80 and 80\,--\,110, respectively as compared to the expected data if it were created from a power law.}

\end{figure}
\begin{figure}
\begin{centering}
\includegraphics[width=0.99\columnwidth]{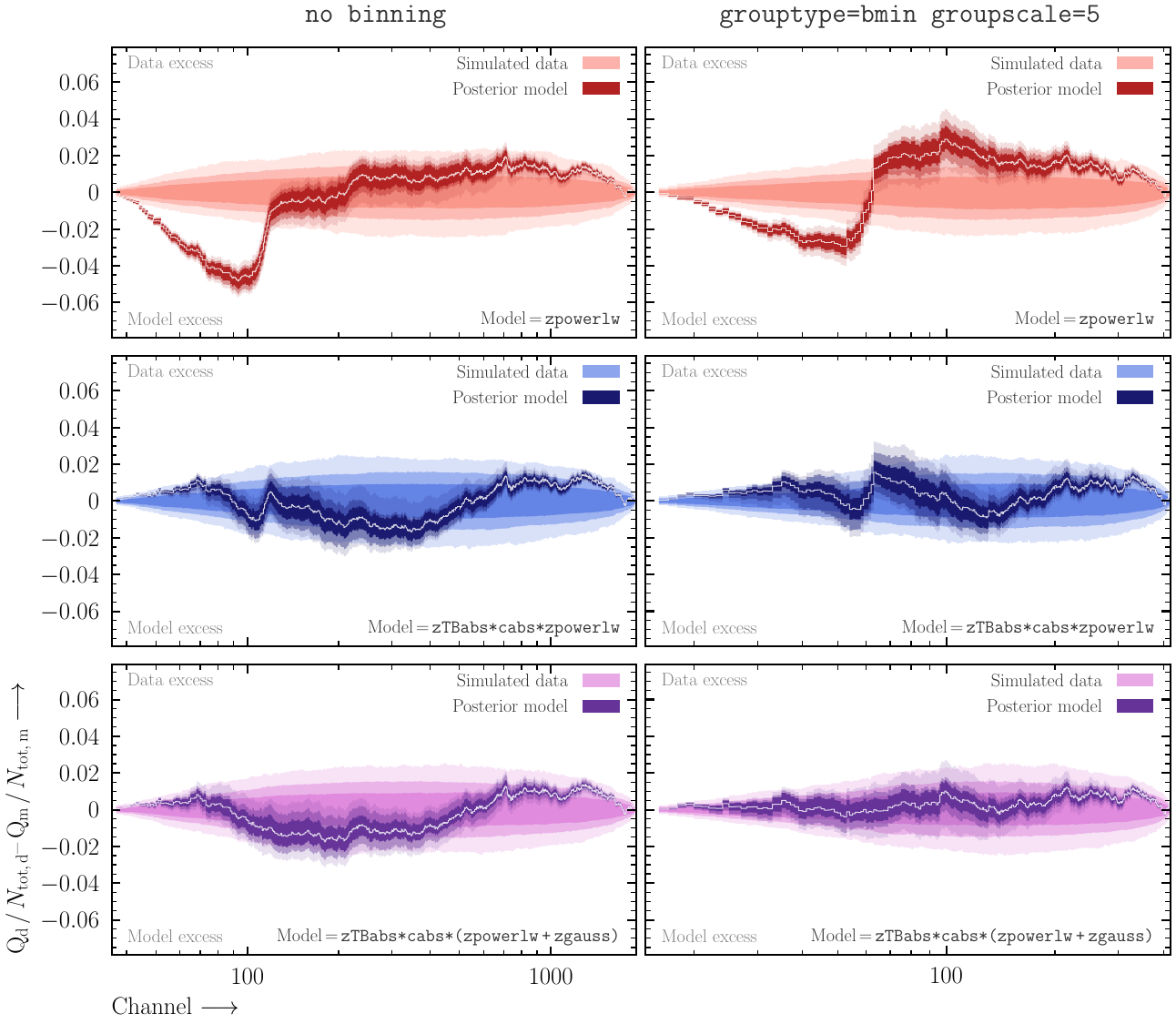}
\par\end{centering}
\caption{\label{fig:qqdiff}\textbf{Q\,--\,Q difference $+$ posterior predictive checks.} In each row, a different model is considered in the plane of energy vs. cumulative difference between the data and model counts (normalised by their totals, respectively). We consider two types of binning: no binning and background binning with 5 counts per background bin minimum on the left and right columns, respectively. In each panel, the horizontal shaded region shows the range that arises purely from fluctuations in the data, when simulated from the model being considered (lower right of each panel). The thin white line and darker, more irregular region then shows the resulting posterior found when fitting the model to the observed data only. In every panel, the shaded regions are separated by opacity with darker to lighter regions showing the 68\%, 90\% and 99\% interquantile ranges. The over-estimation of the source model flux that arises from using Wstat on data that is insufficiently binned is visible in the unbinned (left column), with all model posterior ranges showing a model excess on average relative to the right column.}
\end{figure}

\begin{boxI}
\subsubsection*{Exercise 12 -- posterior predictive checks}
Using a similar process to Exercises~8.1~\&~8.2, simulate 100 spectra from model~1 and save the spectra. Then overplot the range of simulated spectra with the observed counts (e.g., Figure~\ref{fig:ppc_pow}), as well as on a cumulative-difference plot (e.g., Figure~\ref{fig:qqdiff}).\\
\end{boxI}

\subsection{Model comparison}

Bayes' theorem can be applied once more: If we state the posterior
of model i out of a set of competing models as 
\begin{align}
P(\theta|D,M_{i})=\frac{P(D|\theta,M_{i})P(\theta|M_{i})}{P(D|M_{i})}
\end{align}
then the Bayesian evidence looks like a likelihood $P(D|M_{i})$.
Indeed, it is known as the total marginal likelihood, where all parameters have been marginalised out. Applying Bayes' theorem
once more we have:

\begin{align}
P(M_{i}|D)=\frac{P(D|M_{i})P(M_{i})}{P(D)}
\end{align}
where $P(D)=\sum_{i}P(M_{i}|D)P(M_{i})$ normalises the set of models $\{M_{1},M_2,\ldots\}$. What
this means is that we can compute the relative probability of each
\textit{model}, $P(M_{i}|D)$, given computed model evidence values $Z_{i}=P(D|M_{i})$ and assumed model
priors $P(M_{i})$. If we find $Z_{A}=10^{6} \times Z_{B}$
for example, then almost independent of the model priors, the resulting
posterior model probability for model B will be close to zero, and hence model B is strongly disfavored.

In the case of comparing only two models, one can look at the probability ratios:
\begin{align}
\underbrace{\frac{P(M_A|D)}{P(M_B|D)}}_{\mathrm{posterior~odds}}=\underbrace{\frac{P(M_A)}{P(M_B)}}_{\mathrm{prior~odds}}\times\underbrace{\frac{Z_A}{Z_B}}_{\mathrm{Bayes~factor}}
\end{align}



Bayesian model comparison allows weighing the probability of
two or more models. It considers the entire parameter space, with
weighting according to the prior. This has an interesting effect:
Flexible models produce very diverse predictions that are excluded
by the specific data and thus receive low likelihoods over most of
the parameter space. This reduces the integral $Z$, since that is
an average over the parameter space. Therefore, Bayesian model comparison
prefers simpler models (Occam's razor). Simpler is not defined by
the number of parameters, but by the diversity of predictions of the model.

Bayesian inference gives relative probabilities. It does not make decisions.
If we make a decision, for example, discarding the model with lower probability (e.g., $Z_{A}>Z_{B}$
in a priori equally probable models), we do not know the false positive
rate and false negative rate of this decision process. However, we
can apply the process as described in the frequentist `model
comparison' section (Section~\ref{sec:freqmodcomp}): Given our indicator $Z_{A}/Z_{B}$, simulate
under the null model, identify a threshold corresponding to a desired
false positive rate, and apply this with confidence. This has the
benefit of interpretability of the result in a Bayesian sense and
simultaneously a purity guarantee. The main drawback is that the computation
of $Z$ for several hundred simulated spectra can take a long time.
Nevertheless, \cite{Baronchelli2018} demonstrated such false positive 
and false negative computations with BXA for detecting relativistically 
broadened Fe~K lines.

To overcome computational difficulties, approximation to the evidence integral have been proposed. This includes, in order of decreasing quality, the Laplace approximation, which considers the parameter uncertainties and the BIC (Bayesian information criterion), which only considers the maximum likelihood value. Both approximations break down close to the parameter boundaries, and with non-Gaussian posteriors.

The Akaike Information Criterion (AIC), and its variants (Deviance Information Criterion, Widely Applicable Information Criterion) are not based on Bayesian inference. Instead, they compare models based on the information lost when storing the model parameters instead of the data. Better models (with lower information criterion value), thus more completely describe the data. They are useful especially for comparing auxiliary, ad-hoc, empirical models that do not aim to describe the underlying physical process of interest.

\begin{boxI}
\subsubsection*{Exercise 13 -- null hypothesis significance testing with BXA}
Using the same method as in Exercises~8.1~\&~8.2, calculate the Bayes factor thresholds that correspond to type~I/II errors of 1\% for model~1 vs. model~2, as well as model~2 vs. model~3.
\end{boxI}

\newpage
\section{Parameter distributions of a sample}\label{sec4}

\begin{figure}
\begin{centering}
\includegraphics[width=0.5\columnwidth]{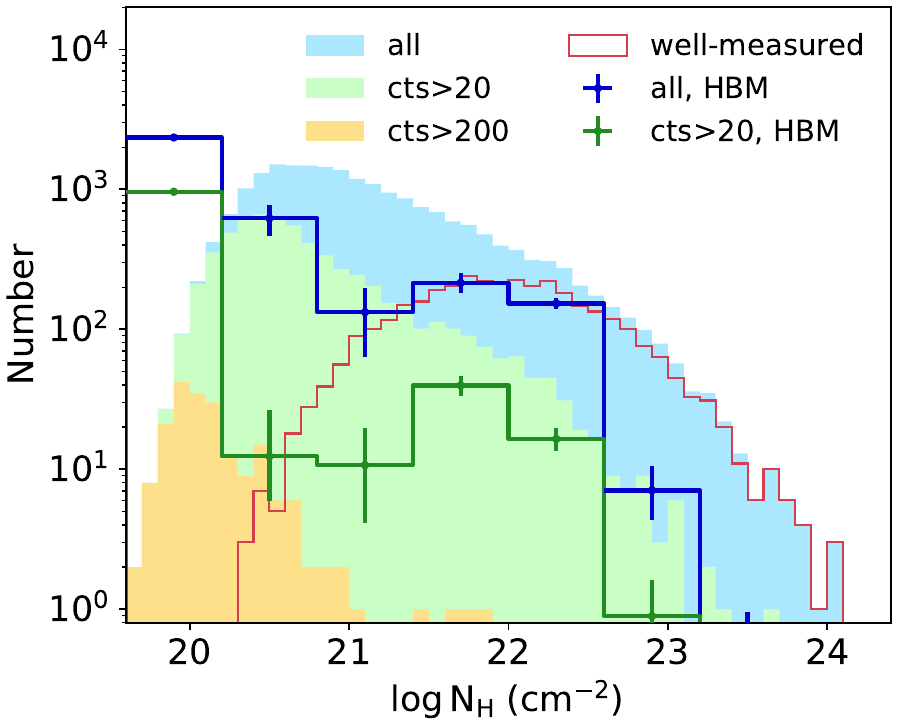}
\par\end{centering}
\caption{\label{fig:samplehist}\textbf{Hierarchical Bayesian Modelling in practice.} The blue filled histogram shows the distribution of the median $N_H$ posteriors measured in 20,000 AGN. Even for sources with $\log N_\mathrm{H}=20$, the median is higher than 20 because of measurement uncertainties extending the posterior from 20 to 22 or 23, depending on the counts obtained. Considering only sources above some source count threshold, the distribution is more concentrated towards 20. The blue step histogram shows the result from the hierarchical Bayesian model (HBM), considering both the sample distribution and the measurement uncertainties of each source. This shows the sample distribution, subject to selection effects, not the distribution of the underlying population.
From \cite{Liu2022eFEDSAGN} with permission from Teng Liu.}
\end{figure}

Finally, let's say you have observed not just the spectrum of one source,
but of several sources from a survey. You notice that each source
has a slightly different $\Gamma$ (each measured with uncertainties).
Now you want to determine the distribution of $\Gamma$ in this
sample. A simple approach here would be to plot a histogram of the
best-fit $\Gamma$ (optima), or the mean or medians of the posterior
PDF. However, this does not consider the uncertainties. This is especially
problematic when the uncertainties are diverse, which is usually the
case.

Diverse uncertainties cause an interesting effect in such histograms.
While the $\hat{\theta}$ estimator from sources with small error
bars will be stable, those with large error bars dilute the histogram
by their scatter. So the histogram shows the intrinsic dispersion of the
sample convolved and broadened by dispersion due to statistical effects.
A real-world example is shown in Figure~\ref{fig:samplehist}. 
We would really like to separate out the measurement effect and 
obtain plausible sample distributions.

That is not trivial with the frequentist approach. In the Bayesian approach,
this is addressed with hierarchical (or multi-level) models.

Let's assume that the true sample distribution of $\theta$ is a Gaussian with mean $\mu$ and standard deviation $\sigma$.
Then we can replace the prior $\pi(\theta)$ with this population
model $G(\theta|\mu,\sigma)$ in eq.~\ref{eq:bayestheorem}. We obtain
an extended posterior which includes the population model parameters
and the per-source parameters $\theta_{i}$:

\begin{equation}
P(\theta_{i},\mu,\sigma|D)\propto\prod_{i}{\cal L}(\theta_{i})\times G(\theta_{i}|\mu,\sigma)\times\pi(\mu,\sigma)
\end{equation}
Now, we are analysing all objects simultaneously. Each object has
free parameters, but at the same time these free parameters are linked
by the Gaussian population distribution. Then, we can sample posteriors
and obtain marginal posterior distributions for $\mu$ and $\sigma$
to learn about the true distribution. This marginalises over the uncertainties
in each object.

How can this be implemented in X-ray spectral fitting packages? Not trivially,
but a numerical approximation is possible. If per-source posteriors
were obtained under flat priors in the parameter of interest, then
the posterior of the population parameters can be approximated as:
\begin{align}
P(\mu,\sigma|D)\propto\prod_{i}\sum_{j}G(\theta_{ij}|\mu,\sigma)\times\pi(\mu,\sigma)
\end{align}
Here, the per-source parameters $\theta_{i}$ have been marginalized
out by averaging over the posterior samples, which are assumed to
be of equal length across the objects. The 
PosteriorStacker\footnote{\url{https://github.com/JohannesBuchner/PosteriorStacker}} package \citep{Baronchelli2018} 
implements this numerical approach for inferring one-dimensional sample
distributions, with a Gaussian population model and a more flexible
histogram-like population model.

\begin{figure}
\begin{centering}
\includegraphics[width=0.9\textwidth]{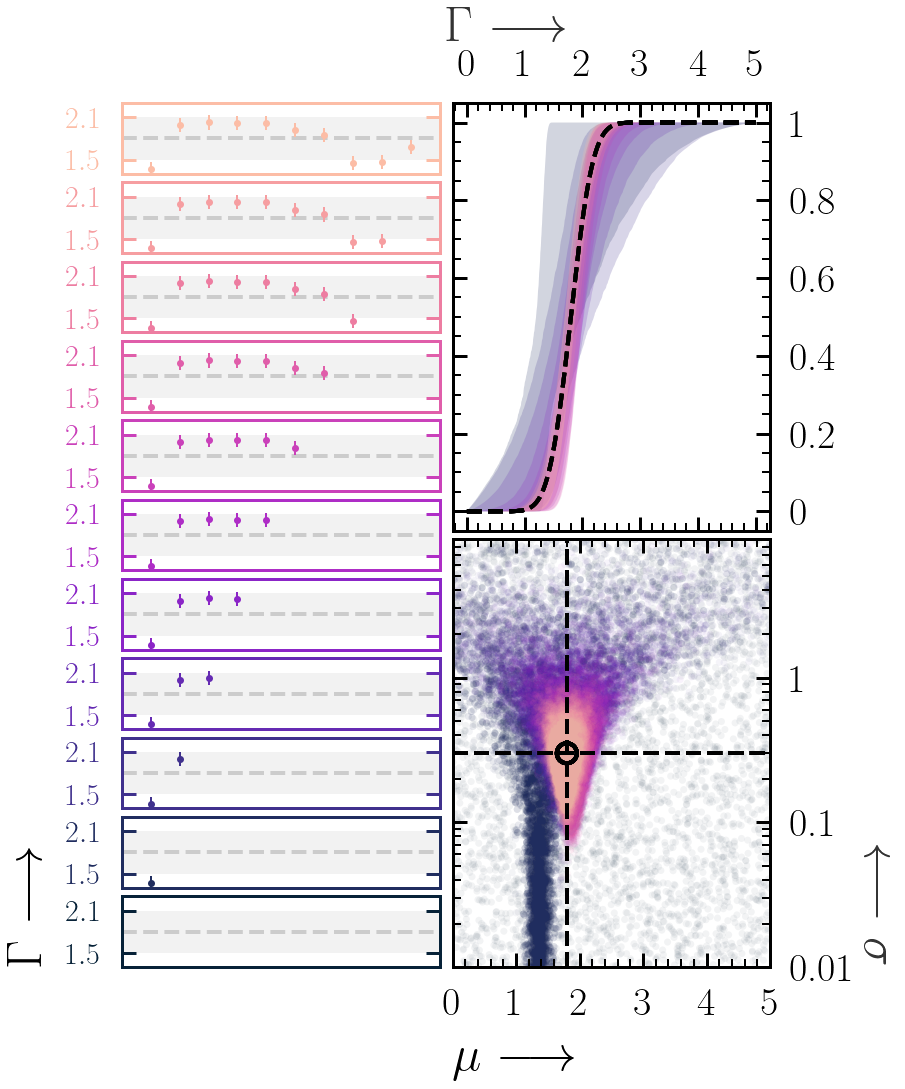}
\par\end{centering}
\caption{\label{fig:hbm}\textbf{Behaviour of single-parameter Hierarchical Bayesian Model.}
A set of 0 to 10 sources is considered (left panels), for each source, a parameter $\Gamma$ was constrained (error bar). We seek to determine the true sample distribution. In this case, it is a Gaussian, with the mean and standard deviation illustrated with the dashed gray line and shade.
For each panel with its color, one PosteriorStacker sample analysis was performed. For the assumed Gaussian sample distribution model, this constrains the mean and standard deviation parameters. Plausible distributions (top right panel), according to posterior samples (bottom right panel), are illustrated with corresponding colors. The truth is shown with a dashed black line.
}
\end{figure}

\begin{boxI}
\subsubsection*{Exercise 14 -- sample distributions}
In the previous exercises, we have estimated the intrinsic photon index for a single source. Here, we want to consider a sample. 
First, simulate a sample of sources from a similar model including a powerlaw, assigning each photon index drawn from a normal distribution. Then, fit each source with the model in BXA, to obtain the posterior samples of photon index for each source. Input these to PosteriorStacker following the manual, and obtain the Gaussian sample distribution. Several examples with different numbers of sources (1 to 10) is shown in Figure~\ref{fig:hbm}.
%
%
%
%
%
\end{boxI}

Inferring population demographics, that is, correcting for selection
effects when the sample was obtained is beyond the scope of this chapter.
For the extension of hierarchical modelling to include sample selection
effects see for example \citep{2004AIPC..735..195L,Loredo2019}. 

\newpage
\section{Further information}\label{sec5}
\subsection*{Books on Statistics}
\begin{itemize}
\item \textquotedbl Introduction to Probability\textquotedbl{} (Dimitri
Bertsekas): covers basics: random variables, combinatorics, derived
distributions, what is a PDF, how to work with them, expectation values 
\item \textquotedbl Probability and Statistics\textquotedbl{} (Morris H.
DeGroot): covers classical approaches to hypothesis testing and frequentist
analysis, so you can really understand, e.g., the classical tests,
what p-values really are, why they are used and how and when they
are useful. 
\item \textquotedbl Scientific Inference\textquotedbl{} (Simon Vaughan):
basic probability theory, statistical thinking, model and data representation
in computers, likelihood function, graphical summaries, basic Monte
Carlo. 
\item \textquotedbl Data Analysis - A Bayesian Tutorial\textquotedbl{}
(Sivia): building an intuition for how Bayesian statistics works 
\item \textquotedbl Bayesian data analysis\textquotedbl{} (Gelman): \url{http://www.stat.columbia.edu/~gelman/book/}
\end{itemize}

\subsection*{Links}
\begin{itemize}
\item A living repository for the figures and exercises used in this Chapter {\small\url{https://github.com/pboorm/xray_spectral_fitting}}
\item `Statistics in XSpec' manual {\small\url{https://heasarc.gsfc.nasa.gov/xanadu/xspec/manual/XSappendixStatistics.html}}
\item Astrostatistics Facebook group: {\small\url{https://www.facebook.com/groups/astro.r/}}
\item XSpec group: {\small\url{https://www.facebook.com/groups/320119452570/}}
\item International CHASC Astro-Statistics Collaboration: {\small\url{http://hea-www.harvard.edu/AstroStat/}}
\item X-ray Spectral Fitting tutorial: {\small\url{http://peterboorman.com/tutorial_bxa.html}}
\end{itemize}
\subsection*{Software packages}
\begin{itemize}
\item Sherpa/CIAO: {\small\url{https://cxc.harvard.edu/sherpa/}}
\item XSpec/HEASoft: {\small\url{https://heasarc.gsfc.nasa.gov/lheasoft/}}
\item BXA (Nested Sampling for XSpec or Sherpa): {\small\url{https://johannesbuchner.github.io/BXA/}}
\item SPEX: {\small\url{https://www.sron.nl/astrophysics-spex}}
\item ISIS: {\small\url{https://space.mit.edu/CXC/ISIS/}}
\item 3ML: {\small\url{https://threeml.readthedocs.io/}}
\end{itemize}

\section*{Acknowledgements}
PB acknowledges financial support from the Czech Science Foundation under Project No.s 19-05599Y \& 22-22643S. PB additionally thanks Vesmír pro Lidstvo for funding the X-ray Spectral Fitting 2022 winter school in Prague.

\bibliographystyle{apalike2}
\bibliography{stats}

\end{document}